December, 2014

# Taxonomic Provenance:
## Two Influential Primate Classifications Logically Aligned

Nico M. Franz[1,*], Naomi M. Pier[1], DeeAnn M. Reeder[2], Mingmin Chen[3], Shizhuo Yu[3], Parisa Kianmajd[3], Shawn Bowers[4], and Bertram Ludäscher[3]

[1] School of Life Sciences, PO Box 874501, Arizona State University, Tempe, AZ 85287, USA
[2] Department of Biology, Bucknell University, 1 Dent Drive, Lewisburg, PA 17837, USA
[3] Department of Computer Science, 2063 Kemper Hall, 1 Shields Avenue, University of California at Davis, Davis, CA 95616, USA
[4] Department of Computer Science, 502 East Boone Avenue, AD Box 26, Gonzaga University, Spokane, WA 99258, USA

* Correspondence to be sent to: School of Life Sciences, PO Box 874501, Arizona State University, Tempe, AZ 85287, USA; E-mail: nico.franz@asu.edu.

Abstract.—Human classifications and phylogenies of perceived natural entities are subject to change in light of new evidence. Taxonomic changes, translated into Code-compliant names and nomenclatural relationships, frequently lead to name/meaning dissociations across succeeding treatments. Classification standards such as the *Mammal Species of the World* (MSW) aim to unify name usages at the global scale, but may nevertheless experience significant levels of taxonomic change from one edition to the next. This circumstance challenges the biodiversity and phylogenetic data communities to develop more granular identifiers to track taxonomic congruence and incongruence in ways that both humans and machines can process, i.e., to logically represent taxonomic provenance across multiple classification hierarchies. Here we show that reasoning over taxonomic provenance is feasible for two classifications of primates corresponding to the second and third MSW editions. Our approach entails three main components: (1) individuation of name usages as taxonomic concepts, (2) articulation of concepts via human-asserted Region Connection Calculus (RCC-5) relationships, and (3) the use of an Answer Set Programming toolkit to infer and visualize logically consistent alignments of these taxonomic input constraints. Our use case entails the Primates sec. Groves (1993; MSW2 – 317 taxonomic concepts; 233 at the species level) and Primates sec. Groves (2005; MSW3 – 483 taxonomic concepts; 376 at the species level). Using 402 concept-to-concept input articulations, the reasoning process yields a single, consistent alignment, and infers 153,111 Maximally Informative Relations that constitute a comprehensive provenance resolution map for every concept pair in the Primates sec. MSW2/MSW3. The entire alignment and various partitions facilitate quantitative analyses of name/meaning dissociation, revealing that approximately one in three paired name usages across treatments is not reliable – in the sense of the same name identifying congruent taxonomic meanings. We assess the feasibility of the RCC-5/reasoning approach and conclude with an optimistic outlook for wider application of logic-based provenance tools in next-generation biodiversity and phylogeny data platforms. [Alignment; Answer Set Programming; classification; concept taxonomy; ontology; Primates; provenance; reasoning.]

*Primatologists on the whole don't understand taxonomy, but they need to, and, in the main, they want to.* (Groves 2001a:vii).

Human classifications of perceived natural groups change in light of new evidence. Over time these changes can affect the validity of taxonomic names and stability of their meanings. Users must keep track of relevant name/meaning relationship updates to communicate reliably about perceived organismal groups and retain the ability to integrate name-annotated information across multiple potentially incongruent classifications. Due to a general trend in biology towards generating and analyzing large synthetic datasets that may reflect heterogeneous taxonomic perspectives, the challenge of reconciling evolving taxonomies is becoming increasingly relevant (Franz et al. 2008).

More than 250 years since Linnaeus' *Systema Naturae,* mammal classifications continue to change at both lower and higher taxonomic levels (e.g., Asher and Helgen 2010; Heller et al. 2013; Zachos et al. 2013; Cotterill et al. 2014). Reasons for such change are manifold, including the application of alternative species concepts or recognition of new phylogenetic information. Although a great number of changes in representing the mammal tree of life are also mirrored in emendations of mammalian nomenclature, the tracking of taxonomic change through the type-centric Linnaean naming system is not perfect (Franz 2005; Kennedy et al. 2005; Franz et al. 2014a; Remsen 2014).

The many-to-many relationships that frequently develop between taxonomic names and their meanings present a challenge to name-based data integration, with at least two apparent solutions. The first is to promote stability and unity in classification, for instance through the adoption of community-wide standards that endorse specific configurations of valid names, synonyms, and taxonomic circumscriptions (Scoble 2004). In the present context, the *Mammal Species of the World* (henceforth: MSW) editions (Honacki et al. 1982; Wilson and Reeder 1993, 2005) are standard references for mammalian classifications that aim to unify name/meaning usages at the global scale. However, even standards experience an evolution of changing taxonomic names and/or meanings from one edition to the next (Patterson 1994; Reeder et al. 2007; Solari and Baker 2007). This is exemplified by the family-level name Cebidae Bonaparte 1831 whose circumscription varies significantly across MSW editions and other treatments (Groves 1993, 2001a, 2005; Rylands and Mittermeier 2009). Standards can mitigate the challenges inherent in name-based data integration – up to a point. They cannot eliminate systemic limitations incurred by using names and nomenclatural relationships as identifiers of exceedingly granular taxonomic incongruences.

The second solution, which may complement the use of classification standards (or, for that matter, of reference phylogenies; see Smith et al. 2014), is to resolve taxonomic similarity and change more precisely than possible with names and nomenclatural relationships alone. In the terminology of computer science, the reconciliation of taxonomic circumscriptions across multiple classifications is a special case of representing (taxonomic) provenance (Cheney et al. 2007). Provenance – herein primarily understood as the tracking of taxonomic information origins and relationships – is an area of growing relevance to information management in biology (Davidson et al. 2007; Zhao et al. 2009). Moreover, provenance can be described formally, and thus inferences of provenance can be enhanced through applications of logic representation and reasoning methods (van Harmelen et al. 2008; Bonatti et al. 2011).

Here we demonstrate that taxonomic provenance is tractable for two incongruent primate classifications – i.e., Groves (1993) and Groves (2005), corresponding to the 2$^{nd}$ and 3$^{rd}$ MSW editions – through a combined approach of taxonomic concept representation and logic reasoning. Our approach has three components. The first of these is to individuate taxonomic name usages through the name sec. author convention introduced by Berendsohn (1995). The sec. (*secundum*) stands for *according to,* and facilitates the use of distinct taxonomic concept labels for (e.g.) Cebidae sec. Groves (1993) versus Cebidae sec. Groves (2005). Properly individuated concepts can be assembled into entire concept hierarchies via parent/child (*is_a*) relationships. This way each separately published perspective – Primates sec. Groves (1993) versus Primates sec. Groves (2005) – can be represented from the ordinal to the species level.

The second component involves asserting an initial, limited set of Region Connection Calculus (RCC-5) articulations (Randell et al. 1992) that express the nature of taxonomic equivalence among concepts pertaining to distinct hierarchies. The available RCC-5 articulations are: congruence (==), proper inclusion (>), inverse proper inclusion (<), overlap (><), and exclusion (|) (Koperski et al. 2000; Franz et al. 2008; Franz and Peet 2009; Weakley 2012; Franz and Cardona-Duque 2013). The input articulations are provided by humans with expertise in the corresponding groups and reflect their understanding of taxonomic provenance in light of the available evidence (e.g., subsumed concepts, homonymy/synonymy, circumscriptions of the phenotype and genotype, distributional data, etc.). For instance, the articulation Cebidae sec. Groves (1993) >< Cebidae sec. Groves (2005) asserts that each concept entails certain congruent subcomponents, yet each also contains additional subcomponents that are unique to it. Uncertainty can be expressed in the form of disjoint articulations (e.g. concept 1 == *or* > concept 2).

Jointly the input taxonomies (T$_1$,T$_2$), articulations (A), and additional constraints (C) that typically apply to taxonomic hierarchies (Thau and Ludäscher 2007) constitute a set of input constraints. The logical consistency of these constraints – i.e., which possible world scenarios exist that satisfy them – can be assessed through a reasoning process, which in turn can produce additional articulations logically implied by the input. The third component of our approach aims to produce a logically consistent, exhaustive, and maximally expressive *alignment* of the input taxonomies. Such an alignment – also called merge taxonomy – amounts to a map of provenance relationships that span across multiple input taxonomies.

Logically aligned taxonomies can inform the integration or separation of data initially linked to only one taxonomy, achieving finer degrees of taxonomic granularity than name-based integration methods. The capabilities and challenges inherent in this approach are shown here for an 800-concept input dataset, the largest to date for which reasoning-based taxonomy alignment has been carried out (Franz et al. 2014a, 2014b).

The Primates sec. Groves (1993 – MSW2) / Primates sec. Groves (2005 – MSW3) alignment use case (henceforth: Prim-UC) is analyzed with the Euler/X software toolkit (Chen et al. 2014a, 2014b). This novel, Open Source toolkit represents and reasons over multi-taxonomy constraints using RCC-5 articulations in combination with Answer Set Programming (Brewka et al. 2011) and custom reasoners. We first describe the toolkit and basic workflow, including input data formats, user/reasoner interactions, and output products. We then characterize the taxonomic input of the Prim-UC, and the conventions used articulate concepts and achieve consistent alignments. Several partitions of the entire 800-concept dataset are made to illustrate different alignment phenomena across taxonomic groups and levels. Concept-level reconciliation of the Prim-UC allows quantitative assessments of name/meaning relation (Franz et al. 2008), which are presented here for the partitioned merges and the entire alignment. The results have implications for provenance tracking of taxonomic concepts used in phylogenetic, comparative, and biodiversity information disciplines. In the Discussion, we suggest pathways to incorporate concept-level representations and alignments into improved provenance practices for annotating and integrating succeeding assembly stages of biodiversity data platforms and the tree of life.

METHODS

*Reasoning Toolkit and Workflow*

The Euler/X software toolkit consists of a set of programming scripts, multiple logic reasoners, and a tree graph visualization system (Thau et al. 2009; Chen et al. 2014a, 2014b; Franz et al. 2014a, 2014b). Given an initial set of input constraints ($T_1, T_2, A, C$), the toolkit delivers the following reasoning and output products (Fig. 1). (1) Visualization of each input taxonomy in the format of an *is_a* hierarchy. (2) Visualization of two input taxonomies and the set of user-asserted articulations (Fig. 1A). (3) Analysis of logical consistency – if the input is not consistent then no alignment is obtained. (4) Reasoning-based diagnosis and removal of jointly inconsistent constraints (constraint overspecification), requiring resubmission of repaired input constraints and return to (3). (5) Inference and representation of one or more consistent alignments, with additional logically implied articulations not provided by the user, in two data representation formats: (a) as the set of Maximally Informative Relations (MIR – interpretable by humans and computers), and (b) as merge taxonomy visualizations (primarily to aid human comprehension) (Fig. 1B). (6) Provision of aggregate views for multiple possible world alignments (constraint underspecification), and decision tree-based reduction of ambiguity in the set of input articulations (leading to more expressive alignments).

A detailed, step-wise account of the input-reasoning-output interaction facilitated by the toolkit is provided in Franz et al. (2014b). In the present contribution we focus on reporting consistent, well-specified, and uniform results for the Prim-UC. We provide all input data files, toolkit commands, and output files (see Supplementary Materials). We have also prepared the entire Prim-UC as an experiment for reproduction at http://recomputation.org/ (Gent 2013). This approach ensures full transparency and permanent accessibility of our data, tools, analyses, and products. To avoid overburdening the narrative with excessive detail on initial, over- or underspecified input configurations, and on concomitant repairs that have led to the consistent and well-specific outcomes, we choose to address these more narrowly targeted issues elsewhere.

*Taxonomic Characteristics of the Primate Use Case*

The Prim-UC is based on two input taxonomies of the MSW2 and MSW3 editions that succeed each other directly, with the latter intentionally referring to the former were appropriate. These taxonomies are highly consistent in their global scope and structure of presentation (Wilson and Reeder 1993, 2005). Incidentally, both are published by the same author, Professor Colin Groves (1993, 2005). Moreover, the majority of changes across editions are rooted in an intermittent reclassification, also authored by Groves (2001a, 2001b). Although the MSW3 edition follows a more information-rich format, both editions provide the following information (where applicable) for each taxonomic concept entry (Fig. 2): valid scientific name; author, year, and citation of the name priority-carrying publication; common name (MSW3 only); type taxon (name) and complete citation; type locality; status of endangerment (CITES); synonyms; and taxonomic comments. The lists of synonymous names are intended to be comprehensive, and the comments are meant to clarify taxonomic perspectives and relationships to other treatments that are either in accordance with the respective edition, or were not adopted for certain reasons.

In spite of the above similarities, the two input taxonomies of the Prim-UC vary significantly in classificatory perspective (Table 1). The taxonomic differences can be divided into several categories; including (1) novel recognition of multiple higher-level ranks for primates in MSW3 (suborder, infraorder, parvorder, superfamily); (2) changes in the mid- to lower-level concept arrangements (family, subfamily, genus); and (3) additions of primate species-level concepts in MSW3, due to either (a) an adherence to more narrowly circumscribed concepts, or (b) the accommodation in MSW3 (2005) of primate phenotypes newly discovered and described *after* MSW2 went to press, i.e., for which there are no taxonomic equivalents in the earlier (1993) perspective. In all, MSW2 recognizes 317 taxonomic concepts, of which 233 correspond to the species level, whereas MSW3 accounts for 483 taxonomic concepts, with 376 at the species level. This means that 86% (143/166) of the differential in the number of primate taxonomic concepts between MSW2 and MSW3 and can be assigned to the later taxonomy recognizing more species-level concepts than the earlier edition.

*Provision of Input Articulations*

The input taxonomies for the Prim-UC may be viewed as compendia that are taxonomically comprehensive and authoritative (Patterson 1994; Solari and Baker 2007). However, each concept entry is treated in an abbreviated form (Fig. 2). Information on diagnostic features or synapomorphic characters that would characterize revisionary and phylogenetic publications is typically omitted. We nevertheless regard the syntactic and semantic content of the taxonomies as sufficiently well specified to permit assertions of concept-to-concept articulations (Figs. 1 and 2). In particular, we provide single 'hybrid' articulations for each concept pair, without distinguishing between intensional (property-referencing) and ostensive (member-referencing) concept components (see Franz and Peet 2009; Franz and Cardona-Duque 2013; Franz et al. 2014b).

The input articulations were asserted with emphasis on the species level (Figs. 1 and 2), and taking into account nomenclatural information, synonymy relationships, lower- and higher-level concept arrangements in each input taxonomy, and additional comments provided. In many instances this information was jointly sufficient to assert unambiguous species-level articulations. Where needed, additional primary literature referenced in Groves (2005) was consulted to resolve articulations (e.g., Groves 2000; Rasoloarison et al. 2000; van Roosmalen et al. 2000).

We stress that assertions of taxonomic concept articulations are not 'objective' (Franz et al. 2014b). Reasoning under the present approach does not represent taxonomic or phylogenetic information directly, but instead proceeds over an asserted RCC-5 translation of such evidence. Different users and representational objectives may therefore produce multiple, alternative alignments for the same input taxonomies. The reliance on user interpretation is an essential part of the workflow. This is not to say, however, that *any* articulation asserted between two concepts is equally well based in evidence. For instance, the Primates sec. Groves (2005) ostensibly refer to (1) *all* taxonomic concepts subsumed under the Primates sec. Groves (1993), *plus* (2) additional taxonomic entities discovered and described after the publication of MSW2. This circumstance instantly eliminates articulations such as inverse proper inclusion (<), overlap (><), or exclusion (|) from the set of justifiable assertions. The remaining options, i.e. Primates sec. Groves (2005) == or > Primates sec. Groves (1993) remain adequate under certain interpretations. Similarly, if all relevant information for two species-level concepts pertaining to the two MSW2/MSW3 taxonomies is relevantly identical (discarding formatting, etc.), and no information that would suggest taxonomic revisions among them is available, then it is unsound to assert any articulation other than congruence (==).

For the purpose of the present alignment, and commensurate with the data presented in the Prim-UC input taxonomies (Fig. 2), we use the following convention. When articulating species-level concepts sec. Groves (2005) to their counterparts sec. Groves (1993), we first determine whether concepts newly recognized at this rank in MSW3 are based (exclusively or overwhelmingly) on *reassessments* of taxonomic boundaries for phenotypic material and/or variability in phenospace *already* considered at the time of publication of the MSW2 compendium. This is apparently the case for 119 of the 143 newly recognized species-level concepts sec. Groves (2005). Examples of such reassessment-contingent species concept additions include *Microcebus griseorufus* Kollmann 1910 sec. Groves (2005) and *Microcebus myoxinus* Peters 1852 sec. Groves (2005). Taxonomic complexities notwithstanding (for review see Rasoloarison et al. 2000), the names that participate in the aforementioned concept labels were originally anchored in phenotypic material that had been recognized (at least implicitly) in MSW2. We therefore represent these articulations as cases of inverse proper inclusion, where union of the more narrowly circumscribed species-level concepts is congruent with the more widely circumscribed concept (Fig. 1). Importantly, the new reassessment-contingent species-level concepts are not considered to expand the circumscriptions of their super-ordinated parent concepts in the respective taxonomies (Fig. 3).

On the other hand, we identified 24 of the 143 additional species-level concepts sec. Groves (2005) as being based (almost) exclusively on phenotypic material that was newly accessioned and evaluated *after* the publication of MSW2 (see also Reeder et al. 2007; Supplementary Materials S4). Because this material was not deemed assignable to any previously established species-level circumscriptions, the 24 new species-level concepts are not readily articulated to any preexisting entities in MSW2. Examples of such accession-contingent additions include *Microcebus sambiranensis* Rasoloarison, Goodman & Ganzhorn 2000 sec. Groves (2005) and *Microcebus tavaratra* Rasoloarison, Goodman & Ganzhorn 2000 sec. Groves (2005). We therefore represent the corresponding articulations as instances of exclusion (Figs. 1 and 3).

Under the herein preferred representation and reasoning approach, additions of species-level concepts rooted in new material accessions (with no previously catalogued phenotypic equivalents) effectively expand the inclusiveness of the super-ordinated MSW3 parent concepts in comparison with their MSW2 counterparts. Because the reasoning approach regards (*ceteris paribus*) the presence of lower-level incongruences as transitive across more inclusive taxonomic ranks (Franz et al. 2014b), we obtain a highest-level articulation of Primates sec. Groves (2005) > Primates sec. Groves (1993). This articulation accounts (minimally) for the historical sequence of human uncovering of primate phenotype diversity, where material grounding 24 species-level concepts subsumed under the Primates sec. Groves (2005) was not recognized some 12 years earlier. The implications of this representation approach are further considered in the Results and Discussion.

*Input/Output Data Formatting, Alignment Partitions, and Toolkit Commands*

All Prim-UC alignments were performed with the Open Source Euler/X software toolkit (Chen et al. 2014a; see also Supplementary Materials), installed on a 4 CPU/ 8 GB RAM Virtual Machine server (http://euler.asu.edu/). Input representation conventions and analyses of alignments are in accordance with Franz et al. (2014b). In particular, we configured and analyzed several partitions of the entire 800-concept alignment, both to demonstrate taxonomically localized and global phenomena of in-/congruence and to produce merge visualizations with taxonomic concept labels that retain legibility under the letter-size space limitations.

We present two sets of six and ten alignment partitions, respectively. The first of these is detailed in Table 2. The six partitions include three alignments that correspond to higher-level concepts in the MSW3 taxonomy sec. Groves (2005) and their MSW2 analogues; viz. (1) the suborder-level concept Strepsirrhini sec. Groves (2005) (with an alignment of 124 x 77 input concepts [MSW3 x MSW2]); (2) the suborder-level concept Haplorrhini sec. Groves (2005), excluding the therein-entailed parvorder-level concept Catarrhini sec. Groves (2005) (169 x 114 concepts); and (3) the parvorder-level concept Catarrhini sec. Groves (2005) (190 x 125 concepts). Additionally, the set includes: (4) an alignment of the entire Prim-UC (483 x 317 concepts); (5) a higher-level subset of that comprehensive alignment, with congruent articulations informed by it, yet limiting the taxonomic rank range to the levels of order to subfamily (38 x 24 concepts); and (6) an alignment of the Hominoidea sec. Groves (2005) and MSW2 analogues (32 x 23 concepts). We show consistent merge visualizations for five of these six alignments in Figs. 4–8. The entire Prim-UC alignment is not shown here due to visualization and legibility constraints.

The second set of Prim-UC partitions entails ten alignments based of the following mid-level concepts sec. Groves (2005) and their MSW2 counterparts: Cheirogaleoidea (see also Fig. 3), Lemuroidea, Lorisiformes, Chiromyiformes, Tarsiiformes, Platyrrhini (excluding Callitrichinae), Callitrichinae, Cercopithecinae, Colobinae, and Hominoidea (repeated). The corresponding merge visualizations are provided in the Supplementary Materials S3.

We adopt the conventions of Franz et al. (2014b) for configurations of Prim-UC input data files (Supplementary Materials S1), output sets of Maximally Informative Relations (MIR; Supplementary Materials S2), and input and merge visualizations. The later MSW3 taxonomy sec. Groves (2005) is consistently represented as $T_2$, whereas the earlier MSW2 taxonomy sec. Groves (1993) is labeled as $T_1$. The $T_2 - T_1$ sequence of annotation is used in all input articulations and output MIR. Similarly, the input and non-congruent merge concepts sec.

Groves (2005) are always illustrated as green rectangles ($T_2$), and the concepts sec. Groves (1993) are shown as yellow octagons ($T_1$) in the visualizations (Fig. 1). Congruent sets of merge concepts are rendered in grey rectangles with rounded corners.

To economize space and maximize consistency between the toolkit input format and visualization products, an abbreviated annotation is used for the taxonomic concept labels, where (e.g.) Primates sec. Groves (2005) becomes "2005.Primates" and *Microcebus murinus* sec. Groves (1993) becomes "1993.Microcebus_murinus" (Fig. 1). Thereby all 800 concepts are unambiguously identified. We utilize this shorthand format in the Results and Discussion.

Two Euler/X toolkit commands were employed to generate the input and output visualizations and alignments. Input data files are annotated with the respective command line arguments. The command `euler -i figure[#].txt --iv` was run to visualize the alignment input (Fig. 1A). The command `euler -i figure[#].txt -e mnpw --rcgo` was used to infer consistent merge alignments under the toolkit's polynomial constraint encoding option (`-e mnpw`). This command simultaneously yields MIR outputs and reduced containment graphs that show overlapping articulations (`--rcgo`) (for further detail see Chen et al. 2014a). The ratio of the number of input articulations and output MIR is provided as a measure of information gained through reasoning (Franz et al. 2014b). The original toolkit visualizations, output as .dot files (GraphViz), were minimally edited with OmniGraffle illustration software (http://www.omnigroup.com/) to obtain consistent spatial renderings of concept groups in accordance with the MSW3 taxonomic arrangement.

The alignments were performed with the toolkit's standard set of Answer Set Programming reasoners, with the exception of the entire 800-concept alignment, which was analyzed with a custom-generated RCC-reasoner (Bowers, personal communication). All alignment input files, toolkit scripts, software dependencies, and run commands for the Prim-UC have been submitted for open, permanent access and identical reproduction at http://recomputation.org/ (see Supplementary Materials S5).

*Analysis of Name/Meaning Relations*

The inferred output MIR (.csv format; Supplementary Materials S2) were variously sorted and compared to perform simple name/meaning relation analyses for the Prim-UC (Geoffroy and Berendsohn 2003; Franz et al. 2008, 2014a). In particular, for each partition we recorded the number of MIR representing each of the RCC-5 articulations (Table 3). The quotient of (1) the number of congruent articulations (==) in an alignment and (2) the number of input concepts in the concept-poorer taxonomy ($T_1$; sec. Groves 1993 in all partitions) provides an approximation of the degree of 'relative congruence' between the two taxonomies for each aligned partition (Table 5). If the ratio approaches one-to-one then relative congruence is high, possibly in spite of differences in name usage and degree of lower-level taxonomic resolution.

Focusing on the entire 800-concept alignment, we furthermore resolve name/meaning identity of articulated MSW3/MSW2 concept pairings by shared taxonomic rank (for MSW2 ranks only; see Table 1), based the following categories: (1) taxonomic congruence, same name(s) (symbolized as == / =); (2) taxonomic congruence, different names (== / ≠); (3) taxonomic proper inclusion, same name(s) (> / =); (4) taxonomic inverse proper inclusion, same name(s) (< / =); and (5) taxonomic overlap, same names(s) (>< / =). "Same name(s)" in the present context means: identical name strings as represented in the input data files (which were not configured to include author names and years). Because both input taxonomies are regarded as Code-compliant (see Witteveen 2014), there are no instances of identically named MSW3/MSW2 concept pairings that are taxonomically exclusive of each other ($| / =$).

Lastly, we consider the ($== / =$) category (1) to reflect 'reliable names', whereas the remaining categories (2-5) above entail concept pairings with 'unreliable names'; in the sense that either identical names ($N_2 = N_1$) identify non-congruent concepts ($C_2 [>, <, ><] C_1$), or congruent concepts ($C_2 == C_1$) are labeled with non-identical names ($N_2 \neq N_1$). A 'reliability ratio' for MSW3/MSW2 names is calculated for each of the six main partitions as the quotient of reliable and unreliable names, adjusting the lower value to 1 (Table 5). Thus a one-to-one reliability ratio reflects that half of the name/meaning relations in the five-category set in an alignment are of the ($== / =$) type.

## RESULTS

We will focus the Results and Discussion on methodological outcomes and implications, as opposed to detailed analyses of taxonomic perspectives and decisions. In particular, we refer to external publications for insights into the ongoing debate on the merits of alternative mammalian species concepts whose differential application is abundantly reflected in the Prim-UC (see, e.g., Groves 2001a, 2001b, 2012, 2013; Baker and Bradley 2006; Asher and Helgen 2010; Frankham et al. 2012; Gippoliti and Groves 2012; Heller et al. 2013; Zachos et al. 2013; Zachos and Lovari 2013; Cotterill et al. 2014).

### Characterization of Alignments and Provenance of Taxonomic Incongruence

The conventions and formulations used to configure the two sets of input constraints for the six and ten respective partitions yield a single, well-specified alignment in each case (Figs. 4–8 and S3–1–10). The consistent scope, presentation structure, and linear reference relation across the two input taxonomies contribute to these highly resolved alignment outcomes. However, the high degree of resolution is paired with frequent and heterogeneous occurrences of taxonomic incongruence, as is reflected in the distribution of green rectangles (concepts unique to $T_2$ – Groves 2005) and yellow octagons (concepts unique to $T_1$ – Groves 1993) in the merge visualizations.

As indicated by the differential number of recognized concepts per taxonomic rank (Table 1), much of the provenance of taxonomic incongruence is assignable to higher numbers of species-level concepts in MSW3. Whereas 119/143 such instances ($\sim 83\%$) are reassessment-contingent (i.e., primarily rooted in narrower species concept circumscriptions adopted in MSW3), the remaining 24 species-level concept additions are grounded in newly accessioned specimen material that was not available for the MSW2 compendium and evidently not subsumable under the earlier species-level concepts (see also Supplementary Materials S4). Narrower taxonomic resolution in MSW3 is also evident at the generic level; e.g., in the recognition of monotypic concepts such as *Mirza* sec. Groves (2005), *Prolemur* sec. Groves (2005), *Pseudopotto* sec. Groves (2005), *Oreonax* sec. Groves, *Symphalangus* sec. Groves (2005), and *Bunopithecus* sec. Groves (2005) – all of which are congruent with species-level concepts already recognized in Groves (1993) yet identified therein by other genus-level name/epithet usages. Interestingly, the balance of the MSW3-narrow/MSW2-wide concept pattern is not fully one-sided. Groves (1993) recognizes four species-level concepts that jointly correspond to the two concepts *Aotus azarae* sec. Groves (2005) and *Aotus lemurinus* sec. Groves (2005) (Figs. 6 and S3–6), with concomitant

MSW3 synonymizations of the names *Aotus infulatus* (Kuhl) and *Aotus brumbacki* Hershkovitz, each recognized as valid in Groves (1993).

A higher-level view of the Prim-UC alignment (Fig. 4) reveals that certain instances of lower-level incongruence propagate to super-ordinated ranks. For instance, the overlapping articulation 2005.Microbus >< 1993.Microcebus 'cascades up' to the family rank in the form of 2005.Microcebus >< 1993.Cheirogaleidae (Fig. 3). The species-level provenance of this articulation is also depicted in Fig. 1. In other instances where only reassessment-contingent differences in lower-level taxonomic resolution are present, these differences can integrate up to congruent super-ordinated concepts. Examples of the latter are: 2005.Hylobatidae == 2005.Hylobatidae and 2005.Hominidae == 1993.Hominidae (Fig. 8).

Groves (2005) introduces three additional, higher-level elements of incongruence in relation to Groves (1993). The first of these is manifested in added partitions above the family level; either (1) through concepts such as 2005.Lemuroidea and 2005.Hominoidea which aggregate multiple reciprocally congruent family-level concepts; or (2) through concepts like 2005.Chiromyiformes and 2005.Tarsiiformes that are taxonomically congruent with their immediate respective children 2005.Daubentoniidae and 2005.Tarsiidae. The second category of provenance of taxonomic incongruence is related to the subfamily level. In particular, Groves (1993) recognizes two subfamily-level concepts 1993.Cheirogaleinae and 1993.Phanerinae whose names have no valid status in Groves (2005), or even a nomenclatural synonym. Such potential analogues are simply not present in MSW3. Conversely, Groves (2005) introduces the concept 2005.Saimiriinae for which there is no immediate nomenclatural or taxonomic counterpart in Groves (1993).

The third category for higher-level incongruence concerns more profound taxonomic differences. Most striking in this category are the differential perspectives of entities that Groves (1993) assigns to the concept 1993.Cebidae (Figs. 4 and 9). This single, widely circumscribed family-level concept has overlapping articulations with the concepts 2005.Cebidae and 2005.Pitheciidae, and additionally properly includes the concepts 2005.Aotidae and 2005.Atelidae and 2005.

Overall, the higher-level alignment (Fig. 4) contains 22 overlapping articulations, of which only three are redundantly depicted in the 2005.Haplorrhini merge (Fig. 6) and two are similarly represented in the 2005.Catarrhini merge (Fig.7). The remaining 16 overlapping articulations reflect the addition of the concepts 1993.Primates at the highest level. The relative congruence is lowest for this alignment with 54.2%, whereas the remaining five focal alignments vary between 86–100% for this indicator (Table 5).

Further analysis of incongruence across mid- to lower-level across partitions reveals (expectedly – see Reeder et al. 2007) that certain alignments are more dynamic than others. For instance, concerning the Strepsirrhini sec. Groves (2005), we observe proportionally more changes in the 2005.Cheirogaleoidea alignment (Figs. 3 and S3–1), with a ratio of 6/25 congruent/unique concept regions, than in the 2005.Lemuroidea alignment with a 22/31 ratio (Fig. S3–2). Similarly, the 2005.Colobinae merge (Fig. S3–9), with a 29/50 ratio, is indicative of proportionally greater taxonomic incongruence than the 2005.Hominoidea merge (Figs. 8 and S3–10), with a 16/16 ratio. These differences are partly rooted in unequal numbers of acquisition-contingent species-level concept additions – four in 2005.Cheirogaleoidea versus zero in 2005.Lemuroidea, and two in 2005.Colobinae versus zero in 2005.Hominoidea – but not entirely. They also appear to reflect discrepancies in precision and reliability of taxonomic and phylogenetic insights across the 1993-2005 period of analysis, in particular with regards to certain Malagasy and New World tropical primate lineages.

*Information Gain and Input Sufficiency*

The Prim-UC partitions provide experiential information on the relationship between the number of sufficient user-asserted input articulations and the information gain conferred through the reasoning process (Franz et al. 2014b). These experiences are connected to the relevant taxonomic constraints, which include (globally, in this use case): non-emptiness, sibling disjointness, and coverage, i.e., a parent concept is complete circumscribed by the union of its children (Thau and Ludäscher 2007). Moreover, no alignment in the Prim-UC has disjoint (ambiguously resolved) input or output articulations, and consequently a single possible world solution is obtained for all alignments. Generally speaking, the number of MIR for an alignment is the product of the number of concepts provided in each input taxonomy (Chen 2014).

Under the above conditions, we observe disproportionate gains of information through reasoning with increasing numbers of input concepts (Table 3). In particular, the ratio of input articulations to input concepts is approximately 1:2 across all alignments. The highest ratio (40:62) is present in the higher-level 2005.Primates alignment, and the lowest ratio (23:55) is observed in the 2005.Hominoidea alignment. Whereas the numbers of sufficient input articulations per alignment range from 23 to 402 articulations (factor of ~ 17.5x), the correlated numbers of MIR increase from 736 to 153,111 (factor of ~ 208x). Accordingly, the information gain varies from 22.8x in the most thoroughly constrained higher-level 2005.Primates alignment, to 380.9x in the entire 800-concept alignment. This amounts to more than a ten-fold disproportionate increase in information gain.

Global application of the coverage constraint in the Prim-UC has the effect that well-specified sets of species-level input articulations 'drive' the alignment of the super-ordinated concept sets (Fig. 1A). For instance, the 2005.Cheirogaleoidae alignment (Fig. 3) entails 21 species-level concepts sec. Groves (2005). Of these, three are congruent with same-ranked concepts sec. Groves (1993), 14 represent reassessment-contingent, narrower concepts that jointly align with only four species-level concepts sec. Groves (1993), and four are acquisition-contingent additions. The corresponding 21 species-level articulations from MSW3 concepts to the corresponding MSW2, plus one additional articulation 2005.Cheirogaleoidae > 1993.Cheirogaleidae, are sufficient to yield the single consistent merge taxonomy. This is so in spite of the presence of 11 intermediate-level concepts in the input – nine at the genus level, two at the subfamily level – that require no user-asserted articulations at all. The practice of exhaustively articulating species-level concepts and then asserting one (or few) additional highest-level(s) articulations was sufficient for producing well-specified alignments for all (6 + 10) Prim-UC partitions. In particular, the number of 402 input articulations of the well-specified 800-concept alignment is only 26 articulations 'above' the minimum needed to articulate each of the 376 species-level concepts sec. Groves (2005) at least once. The Primates sec. Groves (2005) entail 483 concepts in total. This means that consistent, unambiguous articulations for approximately 80 of these were inferred solely on the basis of inferences derived through the reasoning process. Such logical dependencies translate into savings of effort and time for the user and can be appropriated to further optimize the user/reasoner interaction.

*Name/Meaning Relations*

Concept-level resolution of taxonomic provenance in the Prim-UC demonstrates frequent name/meaning dissociation across the two MSW editions, exceeding levels that would be sufficient to accommodate the taxonomic differences themselves. Some measure of dissociation necessarily follows from the 143 species-level concept increase in MSW3, which (under Code-compliant practices) typically means that an inverse proper inclusion/same name(s) ($>$ / $=$) relationship is created among homonymous 2005/1993 concept pairs. An example of this circumstance is the articulation 2005.Microcebus_murinus < 1993.Microcebus_murinus (Figs. 1–3). In total, 55 such instances occur in the Prim-UC (Table 3).

The entire 800-concept alignment contains 203 instances of reliable names (Tables 4 and 5), herein understood as 2005/1993 concept pairings that are taxonomically congruent ($C_2 == C_1$) and whose taxonomic names are identical ($N_2 = N_1$). Of these, 151 are manifested at the species level (for 376/233 MSW3/MSW2 concepts) and 44 occur at the genus level (for 69/60 concepts). Only eight additional instances are present above the genus level, in spite of maximally 24 'opportunities' for reliable names at the subfamily, family, and order levels (Table 4). Unreliable names are overwhelmingly of the taxonomic congruence / different names ($== / \neq$) type (19 articulations) or the aforementioned $>$ / $=$ type (62 articulations). Overlap ($><$) across identically ranked concepts with the same name(s) is rare: twice at the genus level (*Microcebus* and *Pygathrix* sec. Groves 2005/1993; Figs. 3 and S3–9), and once at the family level (Cebidae sec. sec. Groves 2005/1993; Fig. 4). Roughly 3/4 of the unreliable names occur at the species level (224 of 300 same-ranked concept pairs) .

Examination of the reliability ratio, herein taken as the ratio of correlated ($== / =$) versus dissociated ($== / \neq$, *not* $== / =$) name/meaning relationships across the two MSW editions, reveals a 2.1:1 reliable/unreliable names relation across the entire Prim-UC (Table 5). Within smaller partitions, the ratio is least favorable for names as identifiers of congruent taxonomic content at higher levels (partition 2 – 1:1.5), and most favorable in the 2005.Haplorrhini alignment (partition 4 – 1.8:1). Interestingly, even for 2005.Hominoidea alignment (partition 6) which contains only reassessment-contingent species-level concept additions, the reliability ratio is no higher than 1:1. Of the 28 concept-to-concept articulations that reflect either taxonomic congruence ($==$) or nomenclatural sameness ($=$), only 14 are of the combined $== / =$ type.

DISCUSSION

*Feasibility, Assumptions, and Alternative Alignments*

The Prim-UC demonstrates that logically consistent alignments of biological taxonomies are feasible, and scalable to at least 800 concepts per use case. The alignment process, grounded in RCC-5 translations of taxonomic concept provenance, can provide both human- and machine-interpretable outputs, either as sets of Maximally Informative Relations (Supplementary Materials S2) or as merge taxonomy visualizations (Figs. 4–8 and S3–1–10). Depending on the degree of input specification and other representation conventions, the RCC-5/reasoning approach can yield additional, logically inferred articulations in numbers that exceed the user-asserted input articulations by two or more orders of magnitude. The reasoning products can track taxonomic identity at more granular levels than possible using just nomenclatural identities and relationships (homonymy, synonymy). At the same time, they facilitate quantitative assessments of name/meaning relations across the aligned input taxonomies. The latter can inform data integration practices for both systematic and external applications in which names are traditionally used to identify taxonomic content (Franz et al. 2008).

In selecting the Prim-UC, our primary motivation was to showcase the feasibility and utility the RCC-5 alignment approach *under particular conditions.* We lay these out transparently in the Methods and Results sections, provide reproducible input/output data, and have created an instance for comprehensive recomputation (Gent 2013). The identical data and analysis infrastructure will allow others to repeat and thoroughly assess all steps of our work (Supplementary Materials S5).

The input articulations for the Prim-UC are directly derived from the content of the two MSW editions (Fig. 2). They count as third-party articulations (Franz and Peet 2009), posited 'after the fact' by humans who were not involved in authoring the respective input classifications. We recognize that this circumstance is pragmatic rather than ideal, diverging in this regard from other alignments we have published (Franz et al. 2014a, 2014b). The most consequential assumption used to generate the Prim-UC alignments concerns the distinction between reassessment- and acquisition-contingent species-level concept additions in the MSW3 edition (Groves 2005). Three related objections may be offered. We describe and assess the merits of each of these in the following paragraphs.

First, it may be objected that the reassessment/acquisition distinction is not sound, because taxonomic revisions almost always include unequal (though typically increasing) sets of specimens. These sets represent mixtures of previously assessed and newly acquired material (Pullan et al. 2000; Dikow and Meier 2004). It is commonplace in revisionary systematics to combine partial resampling of previously analyzed voucher material with newly acquired specimens to produce novel syntheses. Nevertheless, it remains fruitful to differentiate (1) whether new species-level names are fundamentally derived by reinstating preexisting and synonymous names and associated types as valid, or (2) whether the discovery of thus far undescribed phenotypes requires the creation of names and concepts that have no overlap with published species-level entities. The former case can be thought of as creating finer subdivisions of previously charted geno-/phenospace. The latter case, in turn, genuinely expands that space as recognized in our most recent taxonomies. Thus, while the reassessment/acquisition distinction is not the only form to logically account for the 143 additional species-level concepts sec. Groves (2005), it is adequate in principle and conducive to particular representation and reasoning needs.

The second objection is directly related to the first, and concerns the global acceptance of the coverage constraint (Thau and Ludäscher 2007). An alternative way to formulate this concern is to focus on the absence of intensional, property-centered articulations between above species-level concepts (Franz et al. 2008; Franz and Peet 2009; Franz et al. 2014b). Applying the coverage constraint across the full depth of the input taxonomies means that, *ceteris paribus,* only one acquisition-contingent species-level addition in MSW3 will effectively render the Primates sec. Groves (2005) properly inclusive of (>) Primates sec. Groves (1993) (Fig. 4). One However (as one might argue), the inference that somehow the order-level circumscription of 2005.Primates has expanded over that of 1993.Primates – just by virtue of adding newly discovered species-level species – is unintuitive (at least) in the following sense. We can conceive of a counter-factual situation in which Professor Groves had had access to the publication and specimen material pertaining to *Microcebus sambiranensis* sec. Rasoloarison *et al.* (2000) *at an earlier time,* i.e., when he crafted the MSW2 contribution (Groves 1993). Under that counter-factual scenario, we may ask: had the expert examined this taxonomic entity and its associated material, would he have subsumed it under Primates sec. Groves (1993)? Moreover, would he have assigned the circumscribed entity – in correspondence with inferences made in Rasoloarison *et al.* (2000) – to the genus-level concept *Microcebus* sec. Groves (1993)?

We (humans) sense that the first question can be answered affirmatively, even though the current Prim-UC configuration and reasoning approach do not bear this intuition out. Hypothetically speaking, Groves (1993) ought to have recognized that specimens pertaining to *Microcebus sambiranensis* sec. Rasoloarison *et al.* (2000) are to be classified as 'monkeys' (1993.Primates). Quite likely, he would have also considered them to pertain to the mouse-lemurs (1993.Microcebus) as circumscribed in Groves (1993). Indeed, he approved of this generic placement in his MSW3 contribution (Fig. 3). The global constraint that parents are fully defined by children and the notion that higher-level concepts are understood as congruent by virtue of 'something else than just their children' are in obvious conflict. So then, under what conditions is it adequate to assert that 2005.Microcebus_sambinarensis is exclusive of (|) 1993.Primates, and hence 2005.Primates > 1993.Primates? Conversely, when is it justified to assert that 2005.Microcebus_sambinarensis is included in (<) 1993.Primates, and hence 2005.Primates == 1993.Primates?

The former representation convention (2005.Primates > 1993.Primates, etc.) is that which we have selected for the Prim-UC in the present alignment realization (see above). The alternative – i.e., 2005.Primates == 1993.Primates, etc. – is feasible only under an *intensional reading* of articulations between parent-level concepts (Franz et al. 2008; Franz and Peet 2009; Franz et al. 2014b). For instance, we could resolve to circumscribe 1993.Primates in reference to the following putative synapomorphic trait: "tympanic floor fully ossified, petrosal plate major element, it forms anterior, medial, and posterior walls". This trait (concept) was inferred to represent one of several "good synapomorphies" for Primates sec. Shoshani et al. (1996:114 and 131 – character 13, state 3). The character/state interpretation is traceable to earlier treatments by Wible and Martin (1993) or MacPhee (1981). Other inferred traits in Shoshani et al. (1996; e.g. characters 24, 94) have congruent referential extensions and similar legacies of recurrent taxonomic application.

The predictive, forward-looking nature of character-based definitions of higher-level taxonomic concepts is precisely what would allow us to *affirm* the counter-factual question raised above (Franz and Thau 2010). In particular, we may posit that Groves (1993) was aware of, and endorsed, primate synapomorphies as inferred (e.g.) in MacPhee (1981). If we then furthermore recognize that specimens of *Microcebus sambiranensis* sec. Rasoloarison *et al.* (2000) 'match' these synapomorphic traits (i.e., they are observed as present in them), then 2005.Microcebus_sambinarensis < 1993.Primates becomes an appropriate articulation. We could similarly accommodate the other 23 new, acquisition-contingent species-level concepts of MSW3 under an intensionally circumscribed concept of 1993.Primates, resulting in an articulation at the ordinal concept level of 2005.Primates == 1993.Primates.

Intensional encodings of articulations can represent property-centered congruence among above species-level 2005/1993 concepts in spite of incongruent sets of entailed children (Franz et al. 2014b). Such encodings can be pursued locally, e.g. for select paired genus-level concepts, or even globally. Where applied, they have the effect of representing in-/congruence among parent-level *traits,* i.e., either synapomorphic or otherwise diagnostic features. Trait-level assertions tend to more frequently recover elements of higher-level congruence among concept pairs whose children were differentially sampled between treatments. Using intensional articulations also means that lower-level incongruences need not 'cascade up' to higher-levels (Fig. 3), but instead can be resolved at the next higher level where congruent property-centered circumscriptions occur. Hence the coverage constraint appears locally relaxed.

Returning to the second challenge to our Prim-UC alignment, the issue is not that it is impossible to relax the global coverage constraint in order to obtain more property-centric – and likely more congruent – alignments at higher levels. Our primary reason for opting against this input configuration for this is not logical impossibility, but intellectual deference. Neither input taxonomy directly lists the traits defining higher-level concepts (Fig. 2). Asserting such traits when they are not explicitly stated should be the task of an expert speaker with very immediate access to intensional readings of concepts in each classification (Franz and Peet 2009). Indeed, much information *could* be derived from Groves (2001a); for instance the divergent featured-based diagnoses of *Microcebus* sec. Groves (2001a:68–69) versus *Mirza* sec. Groves (2001a:70–71). In the end, we chose to leave this task to others.

The third and last objection might take issue with the high degree of resolution in the Prim-UC alignments. One might argue that this resolution exceeds the precision of the information provided in the input classifications. Indeed, strict application of the reassessment/acquisition distinction has produced 376 unambiguous species-level articulations, leading to unique merge taxonomies for each input partition (Figs. 3–8 and S3–1–10). This is despite the fact that many currently recognized primate groups have complex taxonomic histories; for instance *Cheirogaleus* (Groves 2000, 2001a) or *Microcebus* (Tattersall 1982; Groves 2001a). Perhaps no unambiguous articulations are appropriate in such cases. More generally, if the source material is *inherently ambiguous* then the alignment realization should not obscure these conditions (for discussion see Franz et al. 2014b).

In summary, each of the aforementioned objections to our realized alignments is potentially meritorious. It is certainly possible that our Prim-UC input specification entails cases of erroneous or overly precise RCC-5 articulations. In each case, the appropriate response is to propose, and taxonomically defend, alternative alignments. Input articulations can be modulated while retaining logical consistency. Coverage constraints can be locally relaxed, and intensional articulations asserted where appropriate. Ambiguity can be expressed through disjoint articulations. These issues are dynamic and represent the proper domain of taxonomic discourse. We predict that the RCC-5/reasoning approach can fully accommodate plausible alternative alignments for the Prim-UC, leading to more nuanced understandings of taxonomic provenance in this and other use cases.

### Concept Provenance-Based Resolution of Taxonomically Annotated Data

What are the advantages of representing taxonomic concept provenance across succeeding classifications and phylogenies with logically inferred RCC-5 articulations? To address this question succinctly, we outline two scenarios of utilizing taxonomic provenance in biodiversity data and phylogenetic information platforms. Additional discussions are developed in Berendsohn (1995), Geoffroy and Berendsohn (2003), Franz (2005), Kennedy et al. (2005), Franz et al. (2008, 2014a, 2014b), Laurenne et al. (2014), Lepage et al. (2014), and Remsen (2014). We note, however, that RCC-5 articulations are not presently tractable with either the well-established Darwin Core biodiversity data or the NeXML phylogenetic data exchange standards (Vos et al. 2012; Wieczorek et al. 2012; Baskauf and Webb 2014).

*Biodiversity data platforms.* Consider using the term "*Microcebus murinus*" to query specimen occurrences documented in the Global Biodiversity Information Facility (GBIF; Edwards 2004).

As of December 2014, this query returns some 560 records whose respective years of acquisition range from 1883 to 2011. Looking at Figs. 1–3, we can immediately identify an opportunity to incorporate provenance with RCC-5 articulations in order to render the query more taxonomically granular and reliable (see also Tables 3–5). In particular, the two MSW standards endorse two non-congruent taxonomic concepts whose respective labels share the name component "*Microcebus murinus*". The earlier 1993.Microcebus_murinus is properly inclusive of (>) the later 2005.Microcebus_murinus.

Suppose that each of the 560 GBIF records was identified explicitly to either the MSW2 or MSW3 primate classification standard. Many biodiversity data platforms allow contributors to specify the classification scheme used to identity specimens (e.g., Constable et al. 2010; Gries et al. 2014). We could then exploit the reasoner-confirmed and -augmented set of logically consistent Prim-UC RCC-5 articulations to formulate queries of the following types (see also Remsen 2014). (1) Return all records identified to the name *Microcebus murinus* (optionally, with synonyms or algorithmically matched names). This query corresponds to the current capability of many portals and services (Patterson et al. 2010; Boyle et al. 2013; Rees 2014b). What follows reaches beyond name-based resolution. (2) Return all records identified to the concept *Microcebus murinus* sec. Groves (1993) and, alternatively, *Microcebus murinus* sec. Groves (2005). Draw the alternative specimen-based distribution maps. (3) Translate all identifications of records to MSW3-endorsed concepts into their corresponding identifications to MSW2-endorsed concepts. This direction of specimen-to-concept identification translation is usually more feasible than the inverse query, because the MSW3 classification is more granular (Table 1). Additional criteria for concept-level resolution such as geographic separation of more narrowly circumscribed entities can aid in achieving reciprocal specimen-to-concept identification translations (Weakley 2012). (4) Highlight 'problem specimens' potentially identifiable to multiple non-congruent concepts, given the set of aligned classification standards that were used (over time) to carry out identifications. (5) Display records in this target region as identified according to the most, or least, granular concept-level taxonomy. (6) For *any* set of specimens or ancillary biological data identified to *any* pair of concepts (there are 153,111 such pairs in the Prim-UC), assess whether the specimens/data can be integrated based on the reasoner-inferred set of MIR. Four out of five articulations – congruence (==), proper inclusion (>), inverse proper inclusion (<), and exclusion (|) – represent direct, unidirectional (>, <) or bidirectional (==, |) provenance information to resolve this query. Most problematic for data integration are overlapping articulations (><, N=49 in the Prim-UC; see Table 3). However, the merge regions that result from overlap are resolvable as such, thereby reducing the challenge to the integration of properly inclusive concepts (Fig. 9).

The above queries and others that leverage taxonomic provenance information via RCC-5 articulations are needed to build more semantically powerful biodiversity data portals (Franz et al. 2008). Although trained humans or Natural Language Processing applications can examine the textual information Fig. 2 and perform many of these tasks 'manually', RRC-5 formalized and logically inferred articulations have advantages over recurrent 'for-human-minds-only' nomenclatural/taxonomic analysis and informal synthesis of provenance (see also Cui 2012). These include: enhanced explicitness, clarity, consistency, and above all machine-interpretability. Ultimately the approach can attain an improved economy of scale (Table 3).

The Prim-UC entails 609 species-level concepts but only 151 instances in which species names reliable identify congruent taxonomic entities across the two MSW editions. Biodiversity data environments and reasoners should have direct, logically actionable access to such information.

As biodiversity science advances deeper into the networked, data-driven age, we should have the ability to perform RCC-5 provenance-based operations on taxonomically annotated data flexibly across platforms, and at scales that humans can no longer effectively process. At the same time, the RCC-5/reasoning approach remains fundamentally dependent on the continuously evolving inferences of taxonomic relationship as advocated by particular human expert speakers, and on reconciliations among multiple succeeding perspectives *as understood by these experts.* Attribution of human expertise is hard-wired into the approach. In short, the RCC-5/reasoning approach leverages both human and computational strengths for synergistic provenance resolution outcomes.

   *Phylogeny assembly platforms.* Reasoning over taxonomic provenance also enhances our understanding of phylogenetically informed concept evolution. Even though the two input classifications of the Prim-UC contain only ranked concepts, the Primates sec. Groves (2005) classification resolves higher-level entities more finely than its predecessor, based on the latest phylogenetic evidence available at the time (reviewed in Groves 2001a). The taxonomic concept approach is equally well suited for aligning informally named ('rankless') clade-level concepts (Franz et al. 2008, 2014b). As discussed above (second objection in the preceding section), assertions of congruence among higher-level concepts are modulated by the choice of applying or relaxing the coverage constraint. If coverage is applied to parent concepts that entail incongruent sets of children due to additions and rearrangements, then overlapping articulations can be frequent among these parents (Fig. 4). This kind of taxonomic overlap is challenging to represent with Linnaean or Phylogenetic nomenclature (Bryant and Cantino 2002; Dubois 2005). However, RCC-5 articulations are ideally suited for this task (Franz and Peet 2009).

   Consider the occurrences of the name "Cebidae" in the alternative MSW2/MSW3 hierarchies (Figs. 4, 6 and 9). The parvorder- to genus-level concept merge depicted in Fig. 9A illustrates the complex articulations of 1993.Cebidae with related MSW3 concepts. In particular, 1993.Cebidae has overlapping articulations ($><$) with 2005.Cebidae and 2005.Pithecidae, due in part to the incongruent assignment of genus-level concepts such as *Callicebus* sec. Groves (1993/2005) or *Callithrix* sec. Groves (1993/2005) to subfamily- and family-level concepts.

   Whenever two *input* concept regions – $C_1$, $C_2$ – are consistently asserted or inferred to overlap, three *output* regions are created in the merge: (1) the merge region that is unique to $C_1$ ($C_1 \backslash C_2$; read: "$C_1$, *not* $C_2$"); (2) the merge region that is unique to $C_2$ ($C_2 \backslash C_1$); and (3) the merge region to which each input region partially contributes ($C_1*C_2$; read: "$C_1$ *and* $C_2$"). Note that none of the three merge regions must necessarily have congruent nomenclatural or taxonomic matches in any of the input taxonomies (Franz et al. 2014a, 2014b). Nevertheless, the ability to refer to these merge regions is critical for achieving consistent, alignment-based data integration.

   The Euler/X toolkit command `euler -i figure[#].txt -e mncb` resolves output regions resulting from overlap according to aforementioned labeling convention (Fig. 9B). The visualization provides valuable insights into the identity of congruent and unique Euler regions in the MSW2/MSW3 merge. In particular, this representation allows users to circumscribe the 1993.Cebidae through sets of subsumed merge regions in which the concepts of Groves (2005) participate. Merge concepts with two parents are of special interest for understanding provenance in this context, because they constitute the shared $C_1*C_2$ regions in cases of overlap. We can thereby understand that the shared regions 2005.Cebidae*1993.Cebidae and 2005.Pithecidae*1993.Cebidae are differentially subsumed under either (1) 2005.Pithecidae and 2005.Cebidae or (2) 1993.Cebidae. The family-level concepts 2005.Aotidae and 2005.Atelidae are also incongruently assigned to MSW2/MSW3 parent concepts (qua unequal ranks). The 2005.Callitrichinae are precisely that region (2005.Cebidae\1993.Cebidae) of the MSW3 classification *not* subsumed under 1993.Cebidae. Meanwhile, 2005.Callithrix contains entities – i.e., seven acquisition-contingent species-level concepts sec. Groves (2005) (see Supplementary Materials S4) – *not* subsumed (2005.Callithrix\1993.Callitrichidae) under 1993.Callitrichidae. An analogous relationship exists for 2005.Callicebus in relation to 1993.Cebidae, i.e., the unique region 2005.Callicebus\1993.Cebidae exists and accommodates three acquisition-contingent species-level concepts sec. Groves (2005). The combination of the reduced containment visualization with overlap (Fig. 9A) and the resolution of that overlap into properly labeled merge concepts (Fig. 9B) provides a maximally granular understanding of provenance for these complex instances of taxonomic incongruence.

Open and dynamic phylogeny assembly platforms (e.g., Smith et al. 2014) stand to benefit from the ability to express taxonomic provenance across alternative or succeeding tree hypotheses (Rees 2014a). Such environments are designed to integrate increasingly well-corroborated inferences of phylogenetic relationship while ensuring exact attribution to intellectual authorship, visualizing mutually supporting and conflicting inferences, and enabling community annotations of evolving phylogenetic content. We have seen above that such directives cannot be built on the basis of taxonomically under-specified names such as "*Microcebus murinus*" or "Cebidae". Instead, all of these services require granular resolution of taxonomic concept identity and provenance as provided by the RCC-5/reasoning approach.

CONCLUSIONS

The Prim-UC demonstrates that it is feasible to represent taxonomic provenance with RCC-5 articulations across incongruent classifications with minimally 400 concepts per input taxonomy. More generally, we have shown that the taxonomic content of classification standards such as the two MSW editions is amenable to logic representation and reasoning. While the Prim-UC entails larger input taxonomies than previous Euler/X toolkit analyses (Franz et al. 2014a, 2014b), the input concepts are explicitly and consistently circumscribed and thus only moderately challenging to articulate. Application or relaxation of the coverage constraint, and optional direct representation of synapomorphic or otherwise diagnostic traits, can yield either member- or property-centric articulations of higher-level concepts. This suggests that the RCC-5/reasoning approach is suited to align a wide range of taxonomic and phylogenetic publications.

Advances in our understanding of primate taxonomy and phylogeny have not stalled since the 3[rd] MSW edition (e.g., Marsh 2014; Pozzi et al. 2014; Rylands and Mittermeier 2014). Many of these advances are sure to find their way into future, heavily used taxonomic standards for this very significant group. In the process, further name/meaning dissociation is bound to occur in comparison to currently circulating standards. The RCC-5/reasoning approach can be used to effectively counteract the loss of taxonomic provenance from the current to the next reference classification.

While the present use case is restricted to aligning taxonomic concepts, the RCC-5/reasoning approach and Euler/X toolkit have wider applicability to systematic or otherwise hierarchically arranged data whose identifiers (names) are subject to semantic drift (Wang et al. 2011). In particular, the approach is also suited to align evolving character concepts and/or differential assignments of specimen vouchers to both concepts and characters (Fig. 10; see also Pullan et al. 2000; Franz 2014). Shifting to such broadly logic-based representations of systematic information will require substantive realignment of virtual data platforms (Kennedy et al. 2005) and data annotation practices (Koperski et al. 2000; Franz 2009; Weakley 2012). At each step, the benefits and costs of more granular name/meaning resolution should be reassessed, and will likely depend on the resolution needs of humans or machines for particular inference tasks. We are optimistic that greater accessibility of logic-based provenance tools will make significant contributions to the integration of systematic information in next-generation data platforms.

ACKNOWLEDGEMENTS

Support of the authors' research through the National Science Foundation is kindly acknowledged (NMF: DEB–1155984, DBI–1342595; BL: IIS–118088, DBI–1147273).

SUPPLEMENTARY MATERIALS

SUPPLEMENTARY MATERIALS S1. – Set of Euler/X toolkit input data files for all alignments produced in the Prim-UC (Figs. 1, 3–10, S3–1–10). Each file is saved is saved in.txt format and contains annotations and instructions for run commands to yield the alignments and visualizations shown in the 19 corresponding figures.

SUPPLEMENTARY MATERIALS S2. – Set of Euler/X toolkit output Maximally Informative Relations (MIR) for the input data files provided in the Supplementary Materials S1. Each output file is saved in .csv format and sorted according to the deduced/inferred output MIR. The MIR files form the basis for analyses of name/meaning relations (Tables 3–5).

SUPPLEMENTARY MATERIALS S3. – Figs. S3–1–3. Alignment visualizations for the set of 10 partitions of the Prim-UC; all most inclusive concepts sec. Groves (2005), with the respective congruent and/or entailed concepts sec. Groves (1993). Fig. S3–1: 2005.Cheirogaleoidea. Fig. S3–2: 2005.Lemuroidea. Fig. S3–3: 2005.Lorisiformes. Fig. S3–4: 2005.Chiromyiformes. Fig. S3–5: 2005.Tarsiiformes. Fig. S3–6: 2005.Platyrrhini (excluding 2005.Callitrichinae). Fig. S3–7: 2005.Callitrichinae. Fig. S3–8: 2005.Cercopithecinae. Fig. S3–9: 2005.Colobinae. Fig. S3–10: 2005.Hominoidea (repeated). See also Supplementary Materials S1 and S3.

SUPPLEMENTARY MATERIALS S4. – List of 24 newly recognized, acquisition-contingent species-level concepts sec. Groves (2005).

SUPPLEMENTARY MATERIALS S5. – The entire Prim-UC has been deposited as an experiment for access and reproduction at http://recomputation.org

REFERENCES

Asher R.J., Helgen K.M. 2010. Nomenclature and placental mammal phylogeny. BMC Evol. Biol. 2010, 10:102.
Baker R.J., Bradley R.D. 2006. Speciation in mammals and the genetic species concept. J. Mammal. 87(4): 643–662.
Baskauf S.J., Webb C.O. 2014. Darwin-SW: Darwin Core-based terms for expressing biodiversity data as RDF. Semantic Web – Interoperability, Usability, Applicability - Special Issue on Semantics for Biodiversity. (In Press).

Berendsohn W.G. 1995. The concept of 'potential taxa' in databases. Taxon 44(2):207–212.

Bonatti P.A., Hogan A., Polleres A., Sauro L. 2011. Robust and scalable Linked Data reasoning Incorporating provenance and trust annotations. J. Web Semant. 9(2):165–201.

Boyle B., Hopkins N., Lu Z., Raygoza Garay J.A., Mozzherin D., Rees T., Matasci N., Narro M.L., Piel W.H., Mckay S.J., Lowry S., Freeland C., Peet R.K., Enquist B.J. 2013. The taxonomic name resolution service: An online tool for automated standardization of plant names. BMC Bioinformatics,14:16, 2013.

Brewka G., Eiter T., Truszczyński M. 2011. Answer Set Programming at a glance. Commun. ACM 54(12):92–103.

Bryant H.N., Cantino P.D. 2002. A review of criticisms of phylogenetic nomenclature: Is taxonomic freedom the fundamental issue? Biol. Rev. 77(1):39–55.

Chen M. 2014. Query optimization and taxonomy integration. Ph.D. Dissertation, University of California at Davis, Davis, CA.

Chen M., Yu S., Franz N., Bowers S., Ludäscher B. 2014a. Euler/X: a toolkit for logic-based taxonomy integration. arXiv:1402.1992 [cs.LO] Available at http://arxiv.org/abs/1402.1992 Accessed December 01, 2014.

Chen M., Yu S., Franz N., Bowers S. Ludäscher B. 2014b. A hybrid diagnosis approach combining Black-Box and White-Box reasoning. In: Bikakis A., Fodor P., Roman D., editors. RuleML 2014. Lect. Notes Comput. Sci. 8620:127–141.

Cheney J., Chiticariu L., Tan W.-C. 2007. Provenance in databases: why, how, and where. Found. Trends. Databases 1(4):379–474.

Constable H., Guralnick R., Wieczorek J., Spencer C., Peterson A.T., The VertNet Steering Committee. 2010. VertNet: A new model for biodiversity data sharing. PLoS Biol. 8(2):e1000309.

Cotterill F.P.D., Taylor P.J., Gippoliti S., Bishop J.M., Groves C.P. 2014. One century of phenetics is enough: Response to "Are there really twice as many bovid species as we thought?" Syst. Biol. 63(5):819–832.

Cui H. 2012. CharaParser for fine-grained semantic annotation of organism morphological descriptions. J. Am. Soc. Inf. Sci. Tec. 63(4):738–754.

Davidson S.B., Boulakia S.C., Eyal A., Ludäscher B., McPhillips T.M., Bowers S., Anand M.K., Freire J. 2007. Provenance in scientific workflow systems. IEEE Data Eng. Bull. 30(4):44–50.

Dikow T., Meier R. 2004. Significance of specimen databases from taxonomic revisions for estimating and mapping the global species diversity of invertebrates and repatriating reliable specimen data. Cons. Biol. 18(2):478–488.

Dubois A. 2005. Proposed rules for the incorporation of nomina of higher-ranked zoological taxa in the International Code of Zoological Nomenclature. 1. Some general questions, concepts and terms of biological nomenclature. Zoosystema 27(2):365–426.

Edwards J.L. 2004. Research and societal benefits of the Global Biodiversity Information Facility. BioScience 54(6):485–486.

Frankham R., Ballouc J.D., Dudash M.R., Eldridge M.D.B., Fenster C.B., Lacy R.C., Mendelson III J.R., Porton I.J., Ralls K., Ryder O.A. 2012. Implications of different species concepts for conserving biodiversity. Biol. Cons. 153:25–31.

Franz N.M. 2005. On the lack of good scientific reasons for the growing phylogeny/classification gap. Cladistics 21(5):495–500.

Franz N.M. 2009. Letter to Linnaeus In Knapp S., Wheeler Q.D., editors. Letters to Linnaeus. Linnean Society of London, London, pp. 63–74.

Franz N.M. 2014. Anatomy of a cladistics analysis. Cladistics 30(3):294–321.

Franz N.M., Cardona-Duque J. 2013. Description of two new species and phylogenetic reassessment of *Perelleschus* Wibmer & O'Brien, 1986 (Coleoptera: Curculionidae), with a complete taxonomic concept history of *Perelleschus* sec. Franz & Cardona-Duque, 2013. Syst. Biodivers. 11(2):209–236.

Franz N.M., Chen M., Yu S., Kianmajd P., Bowers S., Ludäscher B. 2014a. Names are not good enough: Reasoning over taxonomic change in the *Andropogon* complex. Semantic Web – Interoperability, Usability, Applicability - Special Issue on Semantics for Biodiversity. (In Press).

Franz N.M., Chen M., Yu,S., Kianmajd P., Bowers S., Ludäscher B. 2014b. Reasoning over taxonomic change: Exploring alignments for the *Perelleschus* use case. PLoS ONE. (In Press)

Franz N.M., Peet R.K. 2009. Towards a language for mapping relationships among taxonomic concepts. Syst. Biodivers. 7(1):5–20.

Franz N.M., Peet R.K., Weakley A.S. 2008. On the use of taxonomic concepts in support of biodiversity research and taxonomy. In: Wheeler Q.D., editor. The new taxonomy. Systematics Association Special Volume Series 74. Taylor & Francis, Boca Raton, pp. 63–86.

Franz N.M., Thau D. 2010. Biological taxonomy and ontology development: Scope and limitations. Biodiv. Inform. 7(1):45–66.

Gent I.P. 2013. The recomputation manifesto. arXiv:1304.3674 [cs.GL] Available at http://arxiv.org/abs/1304.3674v1 Accessed December 01, 2014.

Geoffroy M., Berendsohn W.G. 2003. The concept problem in taxonomy: Importance, components, approaches. Schrift. Vegetationsk. 39:5–14.

Gippoliti S., Groves C.P. 2012. "Taxonomic inflation" in the historical context of mammalogy and conservation. Hystrix It. J. Mamm. 23(2):8–11.

Gregg J.R. 1954. The language of taxonomy: An application of symbolic logic to the study of classificatory systems. Columbia University Press, New York.

Gries C., Gilbert E.E., Franz N.M. 2014. Symbiota – a virtual platform for creating voucher-based biodiversity information communities. Biodivers. Data J. 2:e1114.

Groves C.P. 1993. Order Primates. In: Wilson D.E., Reeder D.M., editors. Mammal species of the world, a taxonomic and geographic reference, second edition. Smithsonian Institution Press, Washington, D.C., pp. 243–277.

Groves, C.P. 2000. The genus *Cheirogaleus:* Unrecognized biodiversity in dwarf lemurs. Int. J. Primatol. 21(6):943–961.

Groves C.P. 2001a. Primate taxonomy. Smithsonian Institution Press, Washington, D.C.

Groves C. 2001b. Why taxonomic stability is a bad idea, or why are there so few species of primates (or are there?). Evol. Anthr. 10(6):192–198.

Groves C.P. 2005. Order Primates. In: Wilson D.E., Reeder D.M., editors. Mammal species of the world, a taxonomic and geographic reference, third edition. Johns Hopkins University Press, Baltimore, Maryland, pp. 111–184.

Groves C.P. 2012. Species concepts in Primates. Am. J. Primatol. 7(8): 687–691.

Groves C.P. 2013. The nature of species: A rejoinder to Zachos et al. Mamm. Biol. 78(1):7–9.

Heller R., Frandsen P., Lorenzen E.D., Siegismund H.R. 2013. Are there really twice as many bovid species as we thought? Syst. Biol. 62(3):490–493.

Honacki J.H., Kinman K.E., Koeppl J.W. 1982. Mammal species of the world: A taxonomic and geographic reference. Allen Press, Association of Systematics Collections, Lawrence, Kansas.

Kennedy J., Kukla R., Paterson T. 2005. Scientific names are ambiguous as identifiers for biological taxa: their context and definition are required for accurate data integration. In: Ludäscher B., Raschid L., editors. Data integration in the life sciences: Proceedings of the Second International Workshop, San Diego, CA, USA, July 20–22. DILS 2005, LNBI 3615, pp. 80–95

Koperski M., Sauer M., Braun W., Gradstein S.R. 2000. Referenzliste der Moose Deutschlands. Schrift. Vegetationsk. 34:1–519.

Laurenne N., Tuominen J., Saarenmaa H., Hyvönen E. 2014. Making species checklists understandable to machines – a shift from relational databases to ontologies. J. Biomed. Semantics 2014, 5:40.

Lepage D., Vaidya G., Guralnick R. 2014. Avibase – a database system for managing and organizing taxonomic concepts. Zookeys 420:117–135.

MacPhee R.D.E. 1981. Auditory regions of primates and eutherian insectivores. Morphology, ontogeny, and character analysis. Contributions to Primatology, Vol. 18. Karger, Basel.

Marsh L.K. 2014. A taxonomic revision of the Saki Monkeys, *Pithecia* Desmarest, 1804. Neotrop. Primates 21(1):1–163.

Patterson B.D. 1994. Review of "Mammal species of the world: A taxonomic and geographic Reference" by D.E. Wilson; D.M. Reeder. J. Mammal. 75(1):236–239.

Patterson D.J., Cooper J., Kirk P.M., Pyle R.L., Remsen D.P. 2010. Names are key to the big new biology. Trends Ecol. Evol. 25(12):686–691.

Pozzi L., Hodgson J.A., Burrell A.S., Sterner K.N., Raaum R.L., Disotell T.R. 2014. Primate phylogenetic relationships and divergence dates inferred from complete mitochondrial genomes. Mol. Phylogenet. Evol. 75:165–183.

Pullan M.R., Watson M., Kennedy J., Raguenaud C., Hyam R. 2000. The Prometheus Taxonomic Model: A practical approach to representing multiple taxonomies. Taxon 49(1):55–75.

Randell D.A., Cui Z., Cohn A.G. 1992. A spatial logic based on regions and connection. In: Nebel B., Swartout W., Rich C., editors. Proceedings of the Third International Conference on the Principles of Knowledge Representation and Reasoning. Morgan Kaufmann, Los Altos, pp. 165–176.

Rasoloarison R.M., Goodman S.M., Ganzhorn J.U. 2000. Taxonomic revision of mouse lemurs (*Microcebus*) in the western portions of Madagascar. Int. J. Primat. 21(6):963–1019.

Reeder D.M., Helgen K.M., Wilson D.E. 2007. Global trends and biases in new mammal species discoveries. Occ. Pap. Mus. Tex. Tech Univ. 269:1–36.

Rees J.A. 2014a. Thoughts on 'third generation' community taxonomy editing system. Available at https://github.com/OpenTreeOfLife/reference-taxonomy/wiki/Thoughts-on-'third-generation'-community-taxonomy-editing-system. Accessed December 01, 2014.

Rees T. 2014b. Taxamatch, an algorithm for near ('fuzzy') matching of scientific names in taxonomic databases. PLoS ONE 9(9):e107510.

Remsen D. 2014. The use and limits of scientific names in biological informatics. In: Michel E., editor. Anchoring biodiversity information from Sherborn to the 21[st] century and beyond. (In Press)

Rylands A.B., Mittermeier R.A. 2009. The diversity of the New World primates (Platyrrhini): An annotated taxonomy. In: Garber P.A., Estrada A., Bicca-Marques J.C., Heymann E.W., Strier K.B., editors. South American primates, comparative perspectives in the study of behavior, ecology, and conservation, part II. Springer, New York, New York, pp. 23–54.

Rylands A.B., Mittermeier R.A. 2014. Primate taxonomy: Species and conservation. Evol. Anthr. 23(1):8–10.

Scoble M.J. 2004. Unitary or unified taxonomy? Phil. Trans. R. Soc. Lond. B Biol. Sci. 359(1444):699–710.

Shoshani J., Groves C.P., Simons E.L., Gunnell G.F. 1996. Primate phylogeny: Morphological vs molecular results. Mol. Phyl. Evol. 5(1):102–154.

Smith S.A., Cranston K.A., Allman J.F., Brown J.W., Burleigh G., Chaudhary R., Coghill L.M., Crandall K.A., Deng J., Drew B.T., Gazis R., Gude K., Hibbett D.S., Hinchliff C., Katz L.A., Laughinghouse IV H.D., McTavish E.J., Owen C.L., Ree R., Rees J.A., Soltis D.E., Williams T. 2014. Synthesis of phylogeny and taxonomy into a comprehensive tree of life. Proc. Natl. Acad. Sci. (In Press)

Solari S., Baker R.J. 2007. Review of "Mammal species of the world: A taxonomic and geographic reference. 3rd edition." J. Mammal. 88(3):824–830.

Tattersall I. 1982. The Primates of Madagascar. Columbia University Press, New York.

Thau D., Bowers S., Ludäscher B. 2009. Merging sets of taxonomically organized data using concept mappings under uncertainty. In: Proceedings of the 8th International Conference on Ontologies, Databases, and the Applications of Semantics (ODBASE 2009), OTM 2009. Lect. Notes Comput. Sci. 5871:1103–1120.

Thau D., Ludäscher B. 2007. Reasoning about taxonomies in first-order logic. Ecol. Inform. 2(3):195–209.

Van Harmelen F., Lifschitz V., Porter B., editors. 2008. The handbook of knowledge representation. Elsevier, Oxford.

Van Roosmalen M.G.M., van Roosmalen T., Mittermeier R.A., Rylands A.B. 2000. Two new species of marmoset, genus *Callithrix* Erxleben, 1777 (Callitrichidae, Primates), from the Tapajós/Madeira interfluvium, south Central Amazonia, Brazil. Neotrop. Primates 8(1):2–19.

Vos R.A., Balhoff J.P., Caravas J.A., Holder M.T., Lapp H., Maddison W.P., Midford P.E., Priyam A., Sukumaran J., Xia X., Stoltzfus A. 2012. NeXML: Rich, extensible, and verifiable representation of comparative data and metadata. Syst. Biol. 61(4):675–689.

Wang S., Schlobach S., Klein M. 2011. Concept drift and how to identify it. J. Web Semant. 9(3):247–265.

Weakley A.S. 2012. Flora of the southern and mid-atlantic states. Working draft of 30 November 2012. Available at http://www.herbarium.unc.edu/FloraArchives/WeakleyFlora_2012-Nov.pdf. Accessed December 01, 2014.

Wible J.R., Martin J.R. 1993. Ontogeny of the tympanic floor. In: MacPhee R.D.E., editor. Primates and their relatives in phylogenetic perspective. Advances in Primatology Series. Plenum, New York, pp. 111–148.

Wieczorek J., Bloom D., Guralnick, Blum S., Döring M., Giovanni R., Robertson T., Vieglais D. 2012. Darwin Core: An evolving community-developed biodiversity data standard. PLoS ONE 7(1):e29715.

Wilson D.E., Reeder D.M., editors. 1993. Mammal species of the world, a taxonomic and geographic reference, second edition. Smithsonian Institution Press, Washington, D.C.

Wilson D.E., Reeder D.M., editors. 2005. Mammal species of the world, a taxonomic and geographic reference, third edition. Johns Hopkins University Press, Baltimore, Maryland.

Witteveen J. 2014. Naming and contingency: The type method of biological taxonomy. Biol. Phil. (In Press)

Zachos F.E., Apollonio M., Bärmann E.V., Festa-Bianchet M., Göhlich U., Habel J.C., Haring

E., Kruckenhauser L., Lovari S., McDevitt A.D., Pertoldi C., Rössner G.E., Sánchez-Villagra M.R., Scandura M., Suchentrunk F. 2013. Species inflation and taxonomic artefacts – a critical comment on recent trends in mammalian classification. Mammal. Biol. 78(1):1–6.

Zachos F.E., Lovari S. 2013. Taxonomic inflation and the poverty of the Phylogenetic Species Concept – a reply to Gippoliti and Groves. Hystrix It. J. Mamm. 24(2):142–144.

Zhao J., Miles A., Klyne G., Shotton D. 2009. Linked data and provenance in biological data Webs. Brief. Bioinform. 10(2):139–152.

Zimmermann E., Ehresmann P., Zietemann V., Radespiel U., Randrianambinina B., Rakotoarison N. 1997. A new primate species in north-western Madasgar: The golden-brown mouse lemur (*Microcebus ravelobensis*). Primate Eye 63:26–27.

Tables

Table 1. Numbers of taxonomic concepts in the Prim-UC listed per rank, with differentials between the two input taxonomies.

| Taxonomic rank | sec. Groves (1993) | sec. Groves (2005) | Differential |
|---|---|---|---|
| Order | 1 | 1 | - |
| Suborder | - | 2 | + 2 |
| Infraorder | - | 5 | + 5 |
| Parvorder | - | 2 | + 2 |
| Superfamily | - | 4 | + 4 |
| Family | 13 | 15 | + 2 |
| Subfamily | 10 | 9 | − 1 |
| Genus | 60 | 69 | + 9 |
| Species | 233 | 376 | + 143 |
| Total | 317 | 483 | + 166 |

Table 2. Alignment partitions for the Prim-UC, showing the highest-level concept for each input taxonomy and number of entailed taxonomic concepts. Ordered in accordance with taxonomic position and inclusiveness (Fig. 4). See text for further detail.

| Partition | sec. Groves (2005) – MSW3 | Concepts | sec. Groves (1993) – MSW2 | Concepts | Figure |
|---|---|---|---|---|---|
| 1 | Primates | 483 | Primates | 317 | - |
| 2 | Primates – HLO * | 38 | Primates * | 24 | 4 |
| 3 | Strepsirrhini | 124 | Cheirogaleidae, Lemuridae, Megaladapidae, Indridae, Daubentoniidae, Loridae, Galagonidae | 77 | 5 |
| 4 | Haplorrhini ** | 169 | Tarsiidae, Callitrichidae, Cebidae | 114 | 6 |
| 5 | Catarrhini | 190 | Cercopithecidae, Hylobatidae, Hominidae | 125 | 7 |
| 6 | Hominoidea | 32 | Hylobatidae, Hominidae | 23 | 8 |

\* HLO = Higher Levels Only. The range of taxonomic ranks restricted from ordinal to subfamiliar level.
\*\* Excluding Catarrhini sec. Groves (2005).

Table 3. Summary of input concepts, articulations, reasoner-inferred Maximally Informative Relations (MIR), and information gain (MIR/input articulations), for the six Prim-UC partitions, with numbers of specific RCC-5 articulations. See also Table 2.

| Partition | sec. Groves (2005) | Concepts | Articulations | MIR | Gain | == | > | < | >< | \| |
|---|---|---|---|---|---|---|---|---|---|---|
| 1 | Primates | 800 | 402 | 153,111 | 380.9x | 283 | 2053 | 1649 | 49 | 149,077 |
| 2 | Primates – HLO * | 62 | 40 | 912 | 22.8x | 13 | 110 | 27 | 24 | 738 |
| 3 | Strepsirrhini | 201 | 94 | 9548 | 101.6x | 74 | 307 | 246 | 5 | 8916 |
| 4 | Haplorrhini ** | 283 | 150 | 19,266 | 128.4x | 98 | 621 | 423 | 5 | 18,119 |
| 5 | Catarrhini | 315 | 157 | 23,750 | 151.3x | 111 | 558 | 549 | 11 | 22,521 |
| 6 | Hominoidea | 55 | 23 | 736 | 32.0x | 24 | 54 | 63 | 0 | 595 |

\* HLO = Higher Levels Only. The range of taxonomic ranks restricted from ordinal to subfamiliar level.
\*\* Excluding Catarrhini sec. Groves (2005).

TABLE 4. Analysis of taxonomic name/meaning resolution for the Prim-UC (entire 800-concept alignment), categorized by taxonomic rank (for shared MSW2/MSW3 ranks only), and emphasizing entirely or partially taxonomically and nomenclaturally related concepts and names. Legend: $== / =$ : taxonomic congruence, same name(s); $== / \neq$ : taxonomic congruence, different names; $> / =$ : taxonomic proper inclusion, same name(s); $< / =$ : taxonomic inverse proper inclusion, same name(s); $>< / =$ taxonomic overlap, same names(s).

| Rank | sec. Groves (2005) | sec. Groves (1993) | $== / =$ | $== / \neq$ | $> / =$ | $< / =$ | $>< / =$ | Totals |
|------|------|------|------|------|------|------|------|------|
| Species | 376 | 233 | 151 | 17 | 1 | 55 | 0 | 224 |
| Genus | 69 | 60 | 44 | 0 | 7 | 6 | 2 | 59 |
| Subfamily | 9 | 10 | 3 | 0 | 3 | 1 | 0 | 7 |
| Family | 15 | 13 | 5 | 2 | 1 | 0 | 1 | 9 |
| Order | 1 | 1 | 0 | 0 | 1 | 0 | 0 | 1 |
| Totals | 470 | 317 | 203 | 19 | 13 | 62 | 3 | 300 |

TABLE 5. Analysis of relative taxonomic congruence and name reliability in identifying congruent concepts across six Prim-UC partitions (see Table 2). Relative congruence is understood as the quotient of the number of congruent concepts and number of concepts in the concept-poorer taxonomy ($T_1$; sec. Groves 1993). The quotient may be greater than 100% if the concept-richer taxonomy has 'redundant' concepts (multiple superseding ranks with reciprocal taxonomic congruence; see Gregg 1954). Reliable names are of the $== / =$ in Table 4. Unreliable names are of the ($== / \neq$, $> / =$, $< / =$, $>< / =$) types in Table 4. The ratio is reliable names / unreliable names, adjusted to 1 for the smaller value.

| Partition | sec. Groves (2005) | $T_1$ concepts | Actual $==$ articulations | Relative congruence | Reliable names | Unreliable names | Reliability ratio |
|------|------|------|------|------|------|------|------|
| 1 | Primates | 317 | 283 | 89.3% | 203 | 97 | 2.1 : 1 |
| 2 | Primates – HLO * | 24 | 13 | 54.2% | 8 | 12 | 1 : 1.5 |
| 3 | Strepsirrhini | 77 | 74 | 96.1% | 45 | 49 | 1 : 1.1 |
| 4 | Haplorrhini ** | 114 | 98 | 86.0% | 79 | 45 | 1.8 : 1 |
| 5 | Catarrhini | 125 | 111 | 88.8% | 79 | 63 | 1.3 : 1 |
| 6 | Hominoidea | 23 | 24 | 100% | 14 | 14 | 1 : 1 |

* HLO = Higher Levels Only. The range of taxonomic ranks restricted from ordinal to subfamiliar level.
** Excluding Catarrhini sec. Groves (2005).

FIGURE CAPTIONS

FIGURE 1. Illustration of the taxonomic concept representation and reasoning approach. (A) Visualization of input constraints T$_2$ (*Microcebus/Mirza* sec. Groves 2005), T$_1$ (*Microcebus* sec. Groves 1993), and articulations (A) as asserted by the user. Each taxonomic concept hierarchy is separately assembled via parent/child (*is_a*) relationships. The three concepts (*Microcebus griseorufus, Microcebus murinus, Microcebus myoxinus*) sec. Groves (2005) are each properly included (<) in *Microcebus murinus* sec. Groves (1993), based on synonymy information shown in Fig. 2. Four additional species-level concepts sec. Groves (2005) are articulated as exclusive (|) of *Microcebus* sec. Groves (1993) because they are based on phenotypic material for which there was no equivalent in the earlier (1993) edition of MSW2 (Zimmermann et al. 1997; Rasoloarison et al. 2000). The legend indicates the number of nodes and edges for each input taxonomic, and the number of user-asserted input articulations. (B) Visualization (reduced containment graph) of the logically consistent merge taxonomy (alignment) corresponding to the input constraints of (A), showing reasoner-inferred non-/congruent concepts and articulations (see legend). One of two possible world merges is shown. The monotypic genus-level concept *Mirza* sec. Groves (2005) *and* its child *Mirza coquereli* sec. Groves (2005) are taxonomically congruent under the coverage constraint where parent concepts are circumscribed by the union of their children (Thau and Ludäscher 2007). Each is therefore congruent with *Microcebus coquereli* sec. Groves (1993). The two genus-level concepts *Microcebus* sec. Groves (2005) and *Microcebus* sec. Groves (1993) are overlapping; they share two congruent subordinate concepts in the merge while also including reciprocally unique children. The reasoner infers 44 logically implied articulations in the set of Maximally Information Relations (MIR), based on an input of nine initial, user-defined articulations (see also Supplementary Information S1 and S2).

FIGURE 2. Partial representation of primate taxonomic concepts in the 2$^{nd}$ and 3$^{rd}$ editions of the *Mammal Species of the World* series (Wilson and Reeder 1993, 2005). (A) Concept sequence from Primates to *Microcebus* (in part) sec. Groves (1993) (MSW2: 243). (B) Concept sequence from Primates to *Microcebus* (in part) sec. Groves (2005) (MSW3:111–113) (several intermediately ranked concepts omitted). Nine synonymous names are listed under *Microcebus murinus* sec. Groves (1993). Of these, seven are congruently listed under *Microcebus murinus* sec. Groves (2005). However, phenotypes referred to with the names *Microcebus griseorufus* and *Microcebus myoxinus* (listed in bold italics in A) are treated differentially across treatments, acquiring separate species-level concept status from *Microcebus murinus* sec. Groves (2005) in the MSW3 edition. Accordingly, we can assert the articulations (in abbreviated annotation): (1) 2005.*Microcebus_griseorufus* < 1993.*Microcebus_murinus;* (2) 2005.*Microcebus_murinus* < 1993.*Microcebus_murinus;* and (3) 2005.*Microcebus_myoxinus* < 1993.*Microcebus_murinus.* See also Fig. 1.

FIGURE 3. Visualization of the consistent, well-specified Cheirogaleiodae sec. Groves (2005) (T$_2$) / Cheirogaleidae sec. Groves (1993) (T$_1$) alignment; see also Figs. 1 and 2.

FIGURE 4. Visualization of the 2005.Primates – Higher-Levels Only alignment (partition 2; see Table 2), with 22 overlapping articulations of which 16 involve the concept Primates sec. Groves (1993).

FIGURE 5.   Visualization of the 2005.Strepsirrhini alignment (partition 3; see Table 2).

FIGURE 6.   Visualization of the 2005.Haplorrhini alignment, excluding 2005.Catarrhini (partition 4; see Table 2).

FIGURE 7.   Visualization of the 2005.Catarrhini alignment (partition 5; see Table 2).

FIGURE 8.   Visualization of the 2005.Hominoidea alignment (partition 6; see Table 2).

FIGURE 9.   Visualization of the parvorder- to genus-level alignment corresponding to the 2005.Platyrrhini and 1993.Callitrichidae/1993.Cebidae (see also Fig. 6). (A) Reduced containment graph visualization, with overlap (Euler/X command: `-e mnpw --rcgo`). (B) Merge concept visualization (Euler/X command: `-e mncb`), using the $C_1 \backslash C_2$, $C_2 \backslash C_1$, (where "\" = *not*), C1*C2 (where "*" = *and*) convention to identify output merge regions that result from input concept overlap. Recognition of such newly inferred merge regions in the alignment increases the number of inferred articulations (red arrows).

FIGURE 10.   Hypothetical example of an Euler/X toolkit-inferred alignment (1/2 possible worlds shown) whose input entails two incongruent sets of taxonomic concepts, character concepts, and specimen-to-taxonomic/character-concept assignments. The earlier (1990) and later (2010) treatments involve overlapping sets of specimens, where Specimens 1–8 are shared (i.e., re-/examined in each treatment), Specimens 9–10 and 13–14 are only included in the 2010 treatment, and Specimens 11–12 are referenced exclusively in the 1990 treatment. The five species-level taxonomic concepts respectively include 2-5 Specimens and one Property (A–E). Specimens are either directly assigned to parent concepts (as their *is_a* children), or are *identified to* properties (which are like concepts in themselves) via (>, <) articulations. Only two concept-to-concept articulations are provided: (1) 2010.SpeciesName_II == 1990.SpeciesName_II and (2) 2010.GenusName == 1990.GenusName. The reasoning process yields newly inferred articulations among higher-level concepts and their properties, grounded in the specimen-level articulations.

Supplemetary Materials S4. – List of 24 newly recognized, acquisition-contingent species-level concepts sec. Groves (2005). See text for explanation.

1. *Cheirogaleus minusculus* Groves (2000) sec. Groves (2005).

2. *Cheirogaleus ravus* Groves (2000) sec. Groves (2005).

3. *Microcebus berthae* Rasoloarison et al. (2000) sec. Groves (2005).

4. *Microcebus ravelobensis* Rasoloarison et al. (2000) sec. Groves (2005).

5. *Microcebus sambiranensis* Rasoloarison et al. (2000) sec. Groves (2005).

6. *Microcebus tavaratra* Rasoloarison et al. (2000) sec. Groves (2005).

7. *Avahi unicolor* Thalmann and Geissmann (2000) sec. Groves (2005).

8. *Pseudopotto martini* Schwartz (1996) sec. Groves (2005).

9. *Galago rondoensis* Honess (1997) sec. Groves (2005).

10. *Callithrix acariensis* van Roosmalen et al. (2000) sec. Groves (2005).

11. *Callithrix humilis* van Roosmalen et al. (1998) sec. Groves (2005).

12. *Callithrix manicorensis* van Roosmalen et al. (2000) sec. Groves (2005).

13. *Callithrix marcai* Alperin (1993) sec. Groves (2005).

14. *Callithrix mauesi* Mittermeier et al. (1992) sec. Groves (2005).

15. *Callithrix nigriceps* Ferrari and Lopes (1992) sec. Groves (2005).

16. *Callithrix saterei* Silva and Noronha (1998) sec. Groves (2005).

17. *Cebus kaapori* Queiroz (1992) sec. Groves (2005).

18. *Callicebus bernhardi* van Roosmalen et al. (2002) sec. Groves (2005).

19. *Callicebus coimbrai* Kobayashi and Langguth (1999) sec. Groves (2005).

20. *Callicebus stephennashi* van Roosmalen et al. (2002) sec. Groves (2005).

21. *Macaca siberu* Fuentes and Olson (1995) sec. Groves (2005).

22. *Miopithecus ogouensis* Kingdon (1997) sec. Groves (2005).

23. *Pygathrix cinerea* Nadler (1997) sec. Groves (2005).

24. *Trachypithecus ebenus* Brandon-Jones (1995) sec. Groves (2005).

# Fig. 1

A

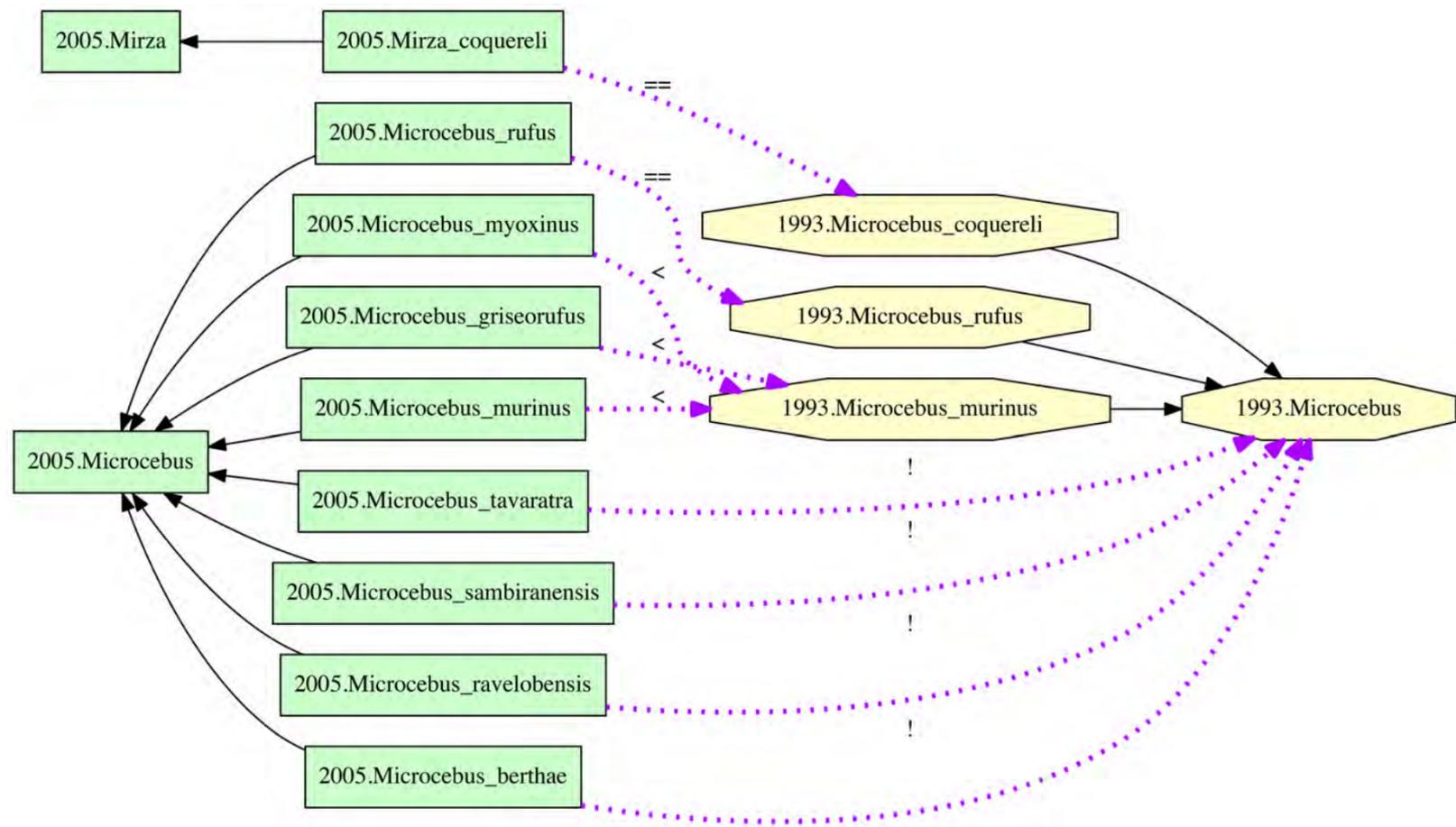

B

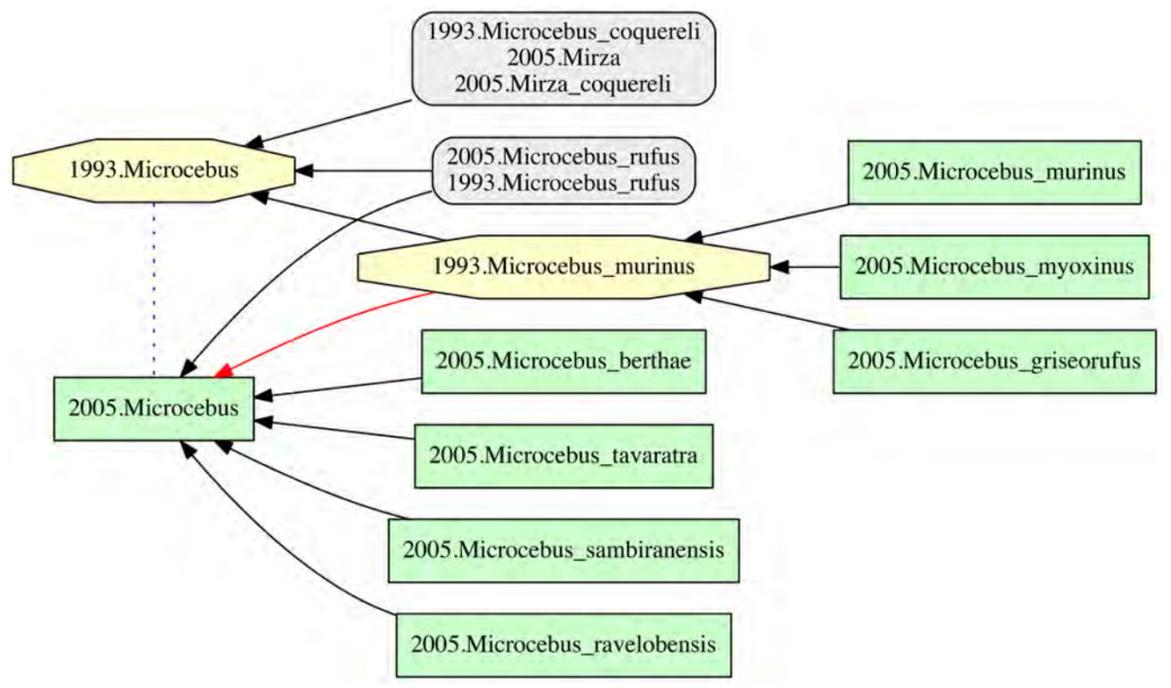

# Fig. 2

**A**

**ORDER PRIMATES**   [sec. Groves 1993]

**Family Cheirogaleidae** Gray, 1873. Proc. Zool. Soc. Lond. 1872:849 [1873].
  COMMENTS: Formerly included in Lemuridae. For status of this taxon, see Rumpler (1875).

**Subfamily Cheirogaleinae** Gray, 1873. Proc. Zool. Soc. Lond. 1872:849 [1873].

***Microcebus*** É. Geoffroy, 1834. Cours Hist. Nat. Mamm., lecon 11, 1828:24.
  TYPE SPECIES: *Lemur pusillus* É. Geoffroy, 1795 (= *Lemur murinus* J. F. Miller, 1777).
  SYNONYMS: *Azema, Gliscebus, Mirza, Murilemur, Myocebus, Myscebus, Scartes.*
  COMMENTS: *Mirza* Gray, 1870, may be a separate genus. Revised by Petter et al. (1977:27-79).

  ***Microcebus murinus*** (J. F. Miller, 1777). Cimelia Physica, p. 25.
    TYPE LOCALITY: Madagascar.
    DISTRIBUTION: W and S Madagascar.
    STATUS: CITES - Appendix I; U.S. ESA - Endangered.
    SYNONYMES: *gliroides,* **griseorufus,** *madagascarensiss, minima, minor,* **myoxinus,** *palmarum, prehensilis, pusillus.*

---

**B**

**ORDER PRIMATES** Linnaeus, 1758.   [sec. Groves 2005]
  COMMENTS: Fully reviewed by Groves (2001*c*), whose arrangement is followed here, with the addition of some subsequently
    desribed species. [...]

[Entries for Suborder Strepsirrhini, Infraorder Lemuriformes, Superfamily Cheirogaleoidea]

**Family Cheirogaleidae** Gray, 1873. Proc. Zool Soc. Lond. 1872:849 [1873].
  COMMENTS: Formerly included in Lemuridae. For status of this taxon, see Rumpler (1975) and Groves (2001*c*), who reviewed and
    rejected the hypothesis that they may be more closely related to (non-Malagasy) Lorisiformes than to (Malagasy) Lemuriformes.

***Microcebus*** É. Geoffroy, 1834. Cours Hist. Nat. Mamm., lecon 11, 1828:24.
  TYPE SPECIES: *Lemur pusillus* É. Geoffroy, 1795 (= *Lemur murinus* J. F. Miller, 1777).
  SYNONYMS: *Azema* Gray, 1870; *Gliscebus* Lesson, 1840; *Murilemur* Gray, 1870; *Myocebus* Wagner, 1841; *Myscebus* Lesson, 1840;
    *Scartes* Swainson, 1835.
  COMMENTS: Revised by Rasoloarison et al. (2000).

  ***Microcebus griseorufus*** Kollman, 1910. Bull. Mus. Histo. Nat. Paris, 16:304.
    [Entries on common name, type locality, distribution, status (CITES)]

  ***Microcebus murinus*** (J. F. Miller, 1777). Cimelia Physica, p. 25.
    [Entries on common name, type locality (fixed by Rasoloarison et al. 2000), distribution, status (CITES)]
    SYNONYMS: *gliroides* A. Grandidier, 1868; *madagascarensis* É. Geoffroy, 1812; *minima* Boddaert, 1785; *minor* Gray, 1842;
      *palmarum* Lesson, 1840; *prehensilis* Kerr, 1792; *pusillus* É. Geoffroy, 1795.

  ***Microcebus myoxinus*** Peters, 1852. Reise nach Mozambique, Zool. 1:14.
    [Entries on common name, type locality (restricted by Rasoloarison et al. 2000), distribution, status (CITES)]

# Fig. 3

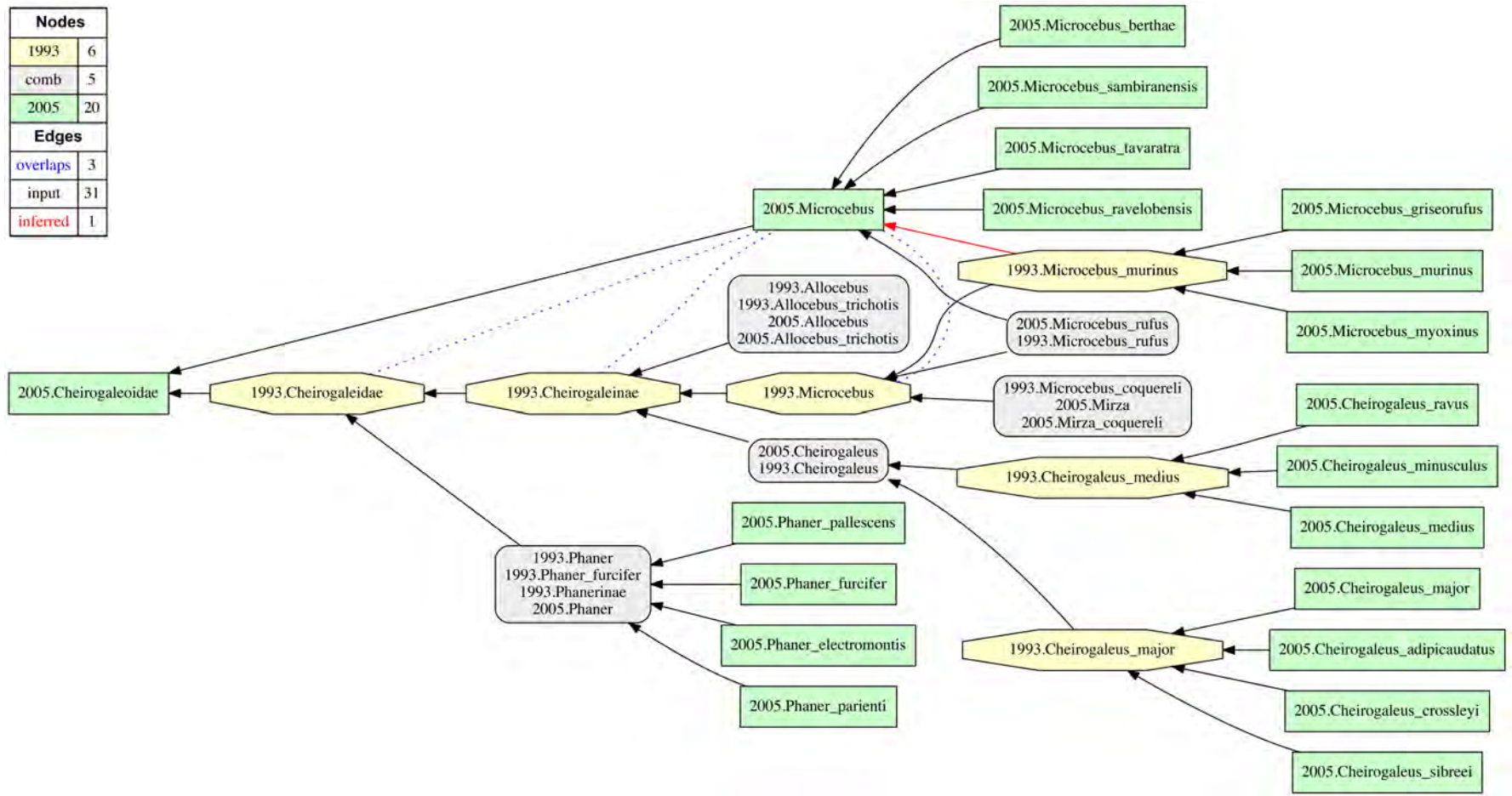

# Fig. 4

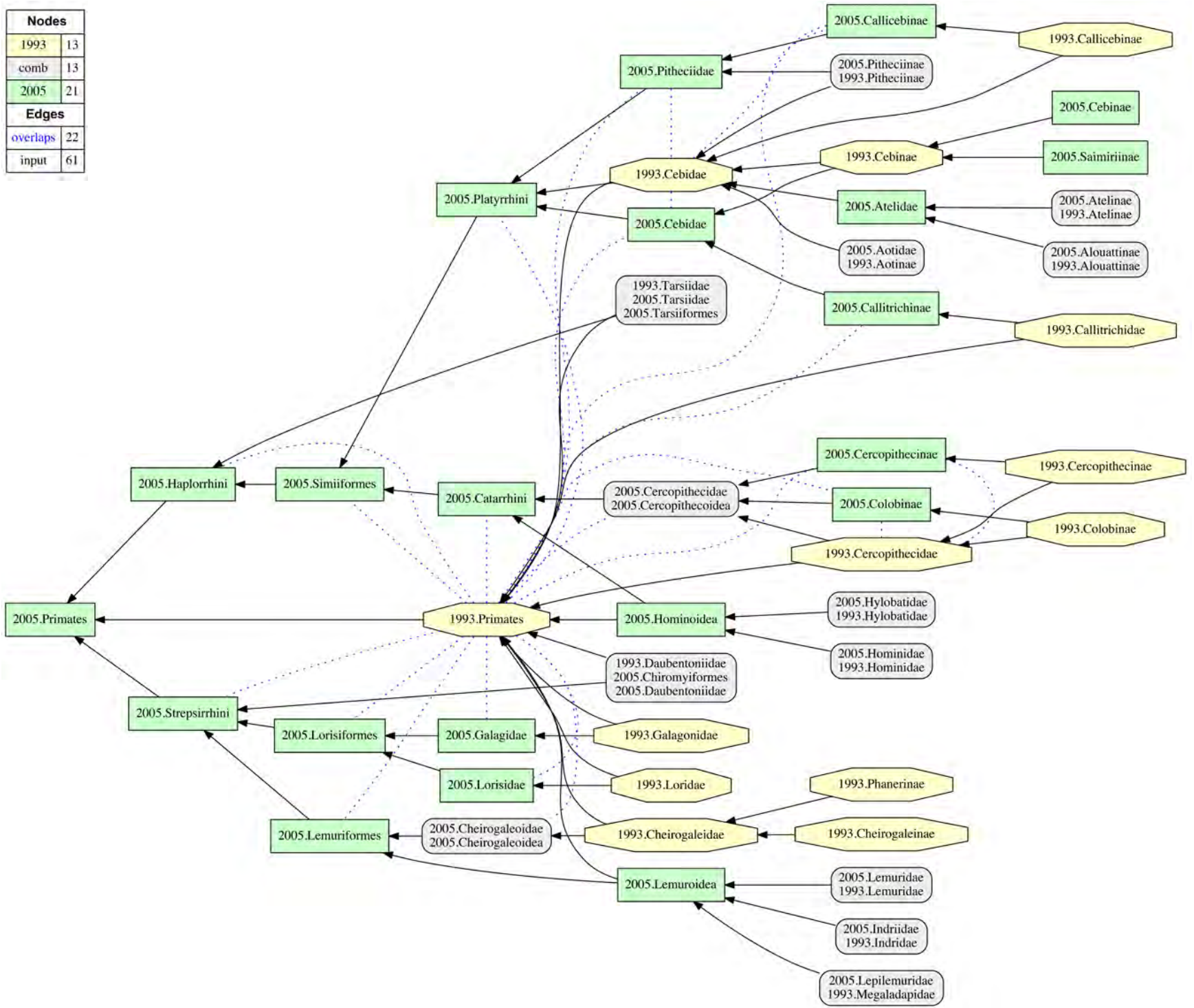

# Fig. 5

# Fig. 6

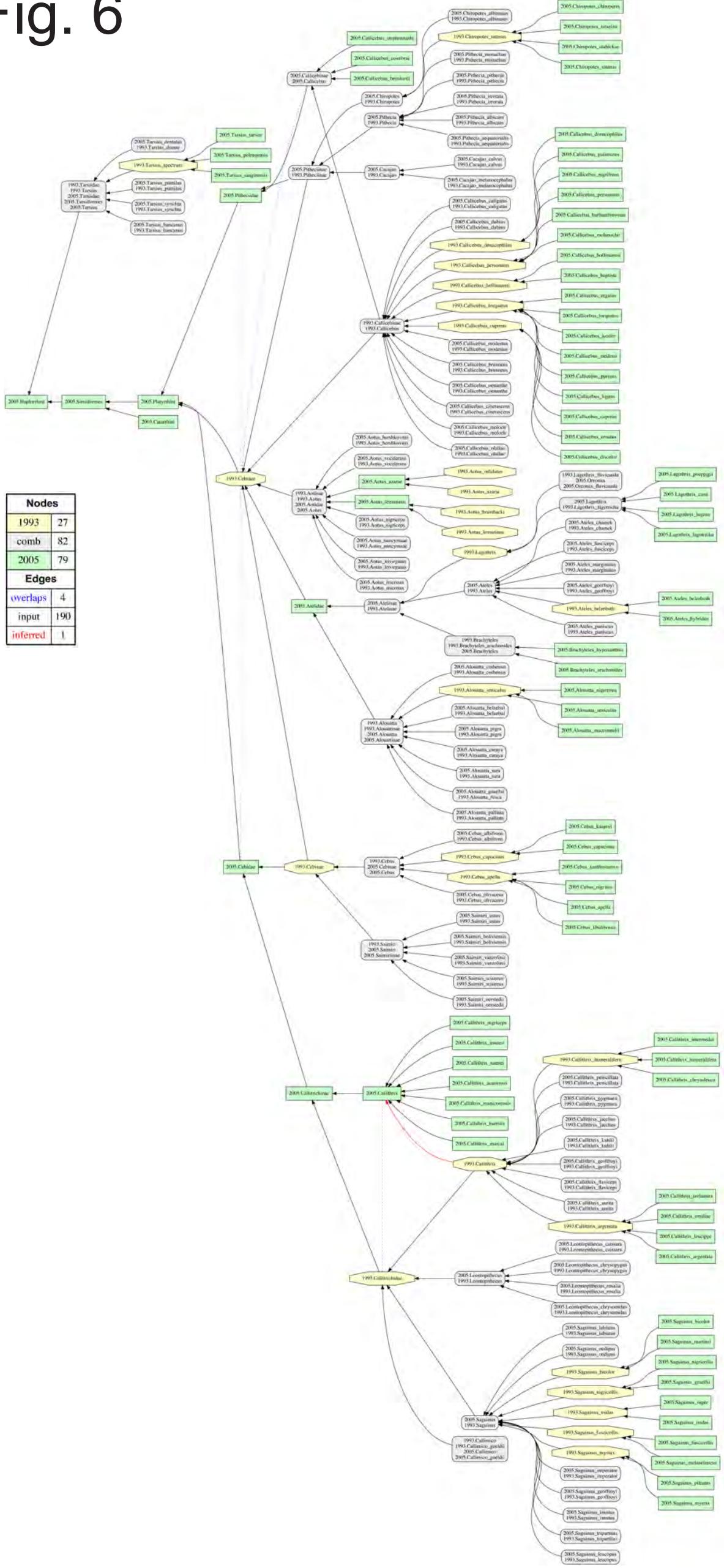

# Fig. 7

# Fig. 8

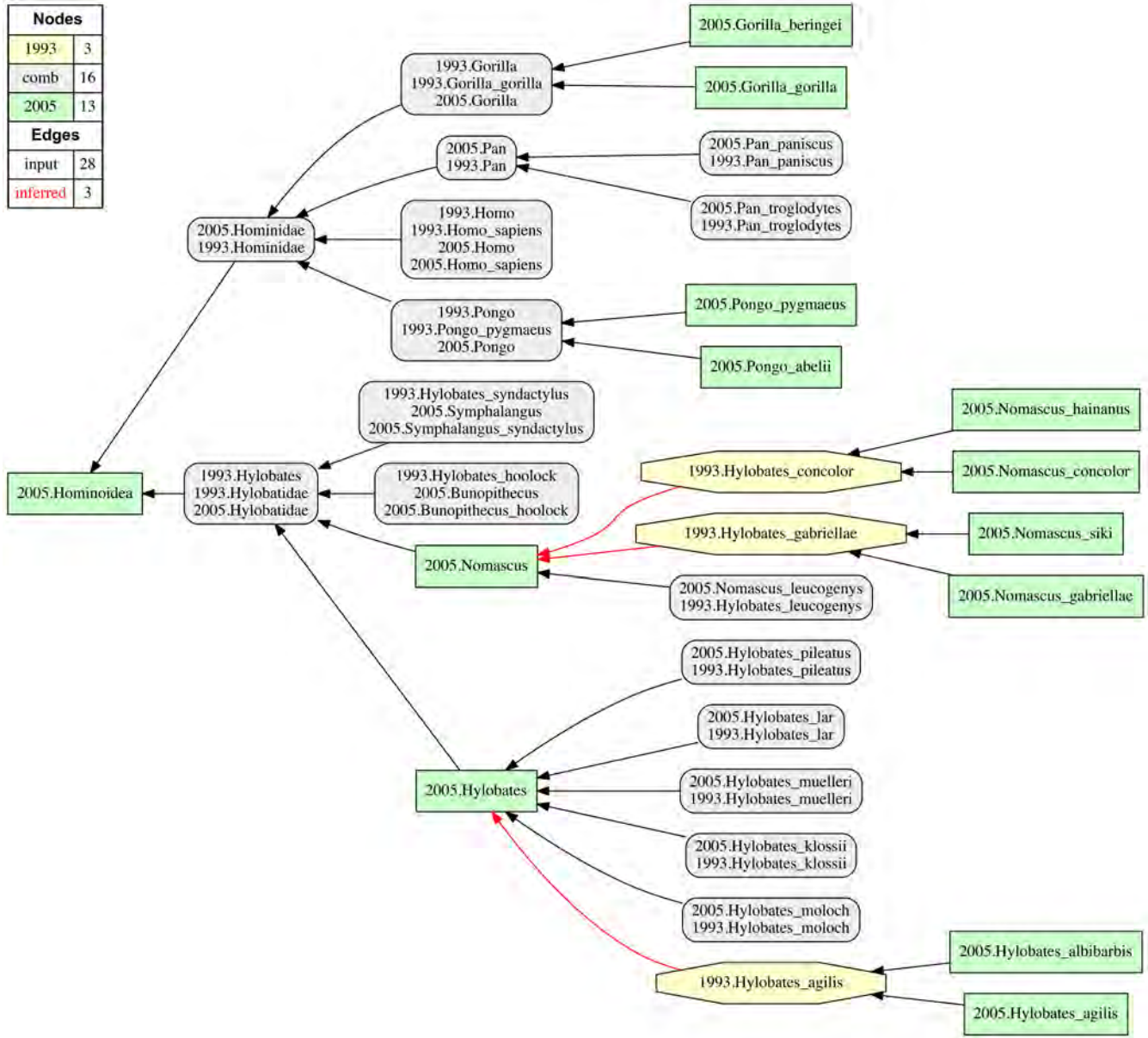

# Fig. 9

# Fig. 10

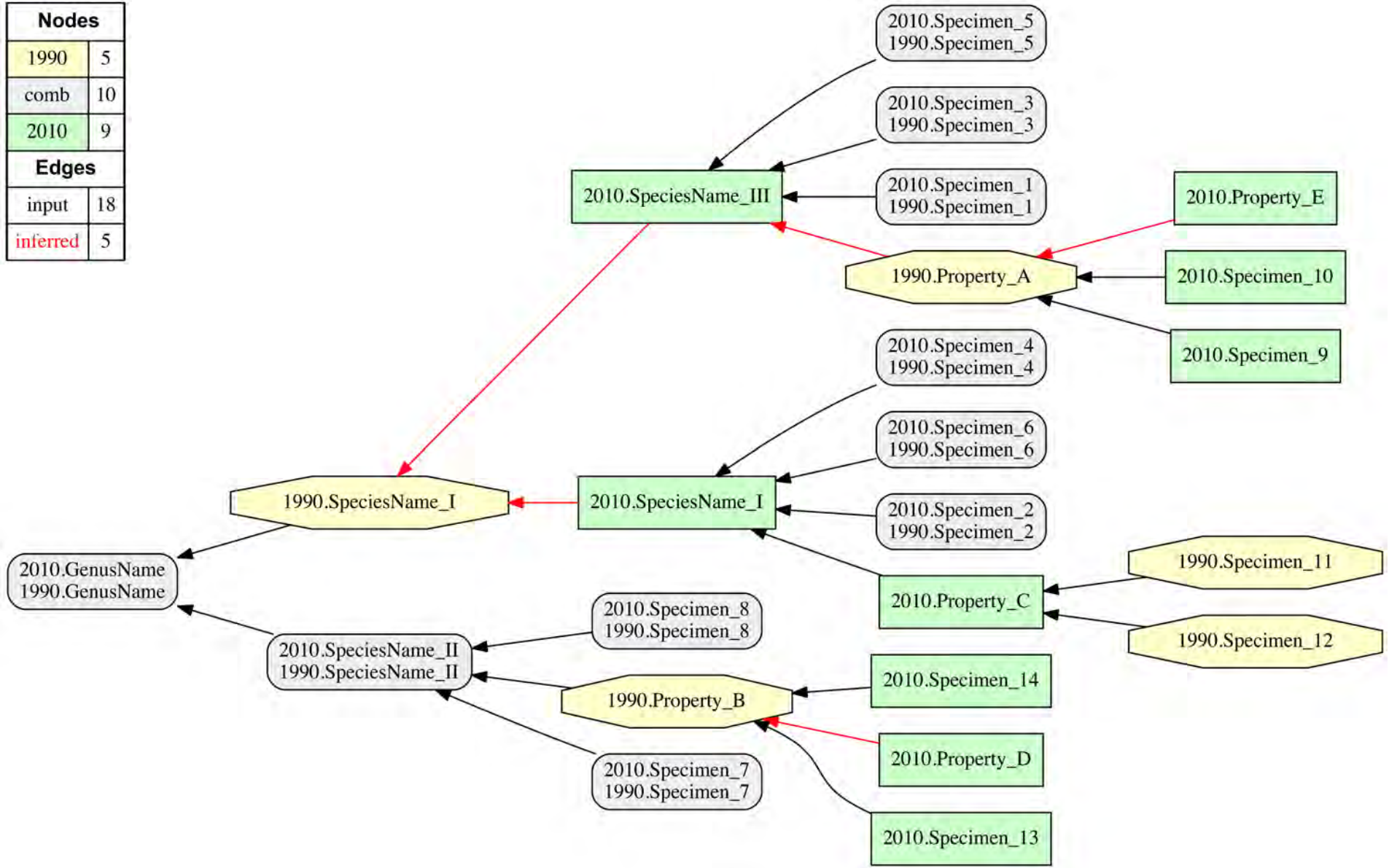

# Fig. S3-1

# Fig. S3-2

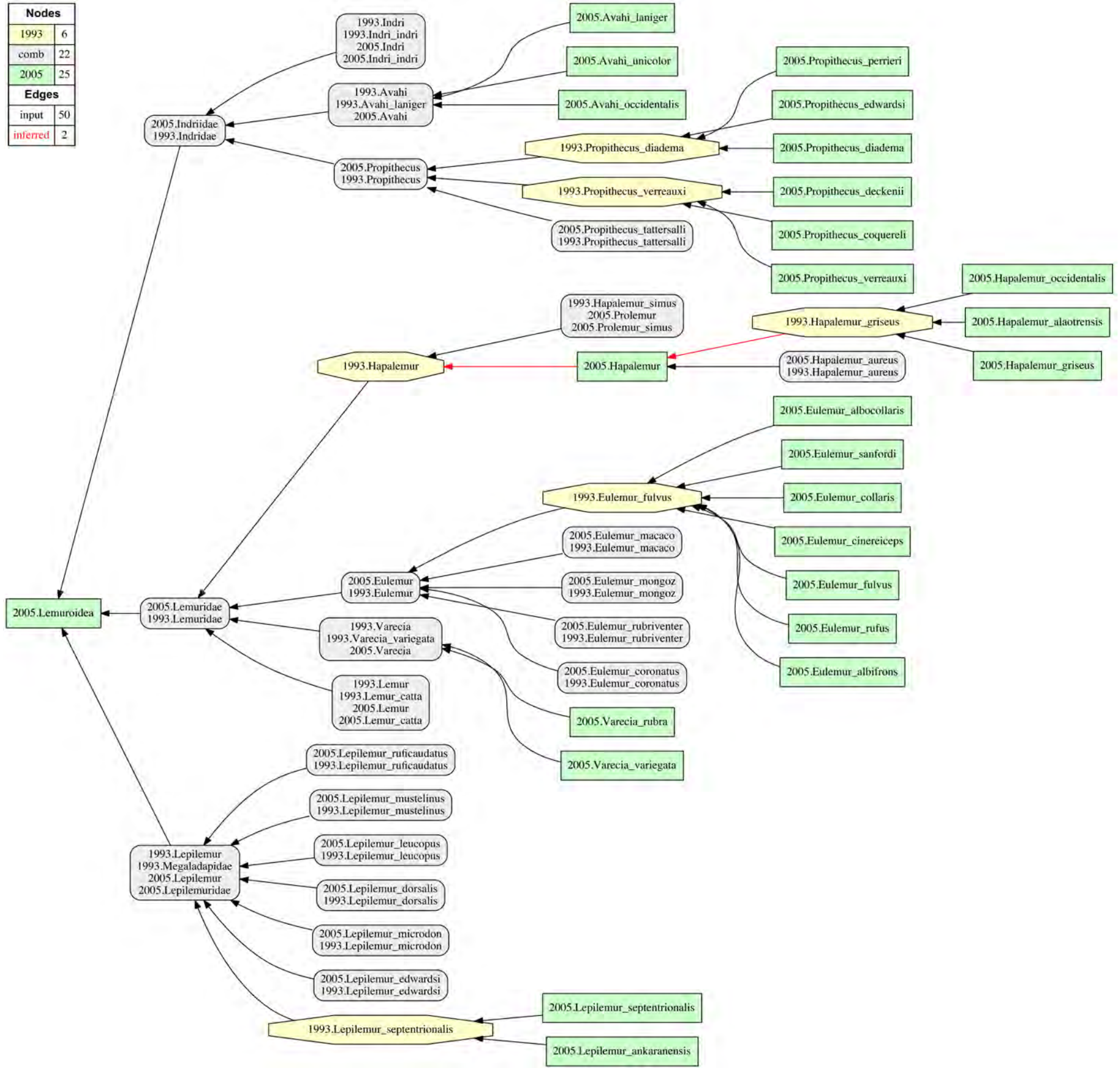

# Fig. S3-3

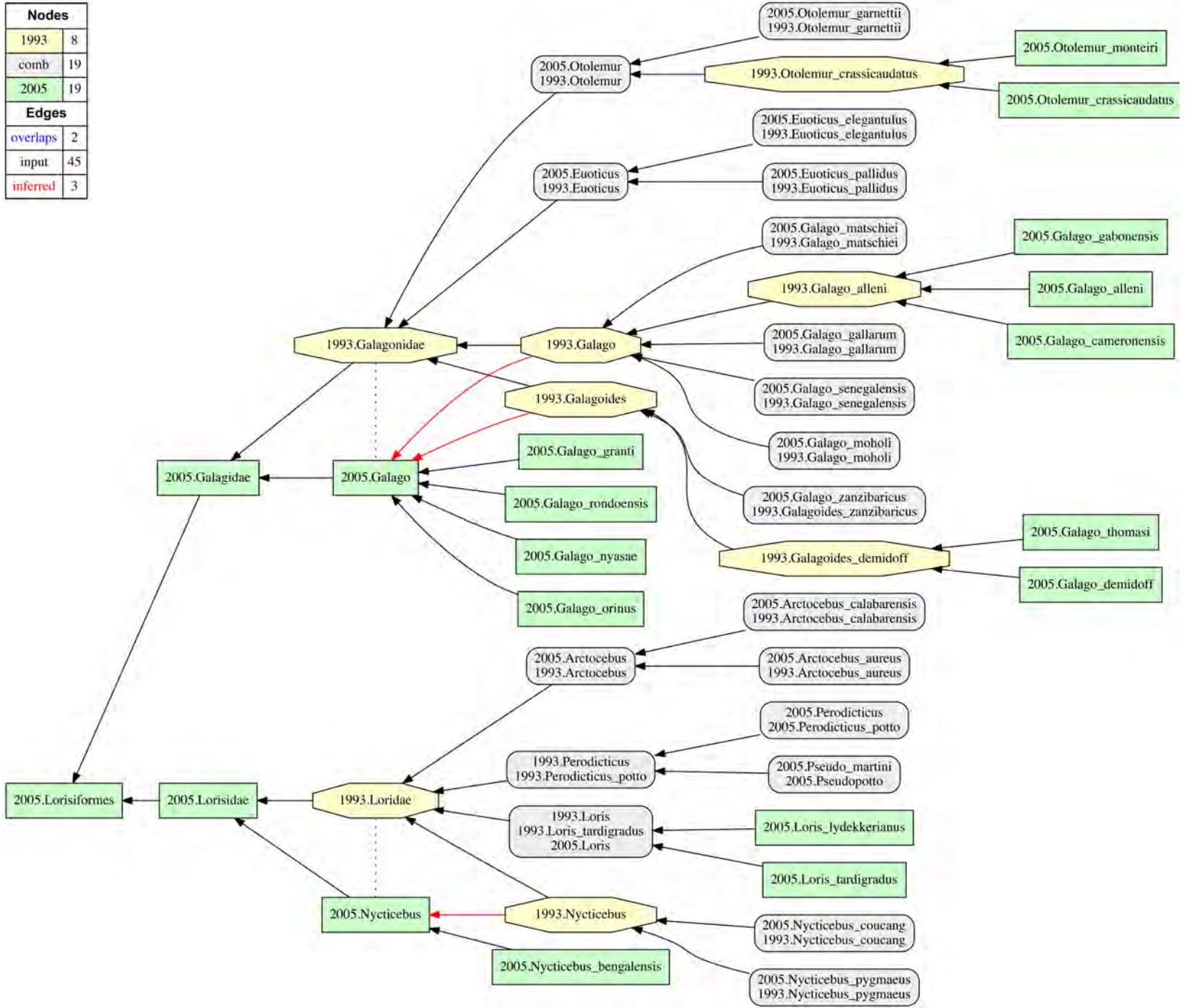

# Fig. S3-4

| Nodes | |
|---|---|
| comb | 1 |
| **Edges** | |

1993.Daubentonia
1993.Daubentonia_madagascariensis
1993.Daubentoniidae
2005.Chiromyiformes
2005.Daubentonia
2005.Daubentonia_madagascariensis
2005.Daubentoniidae

# Fig. S3-5

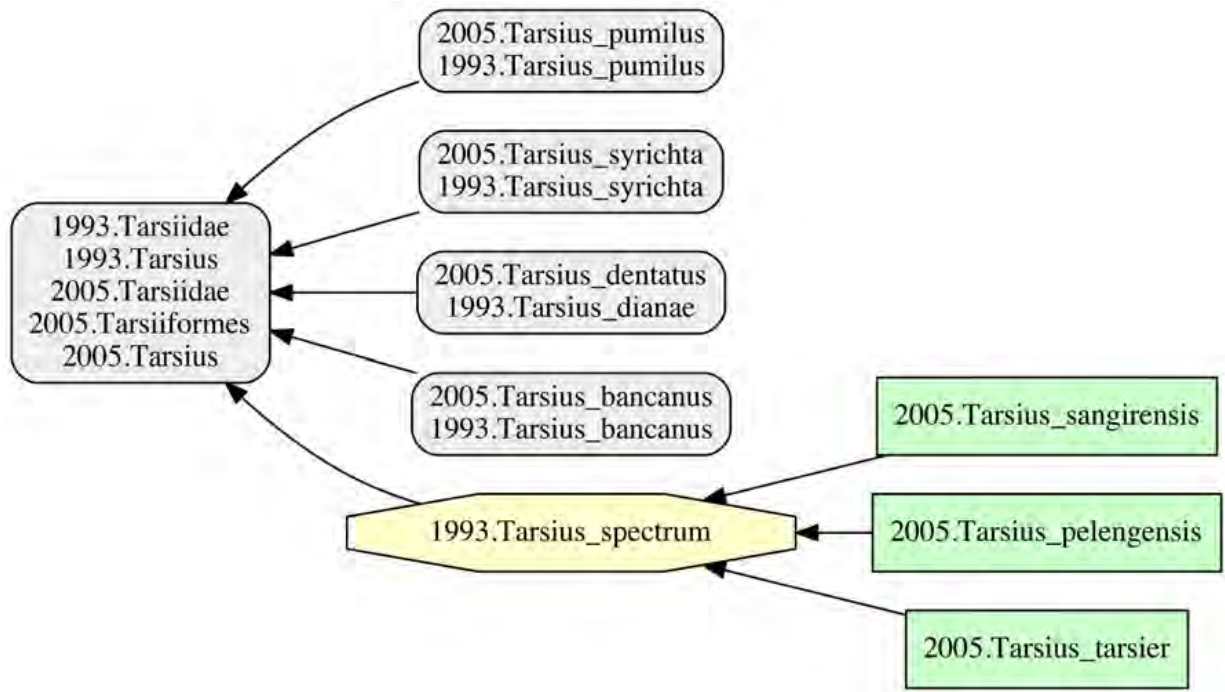

# Fig. S3-6

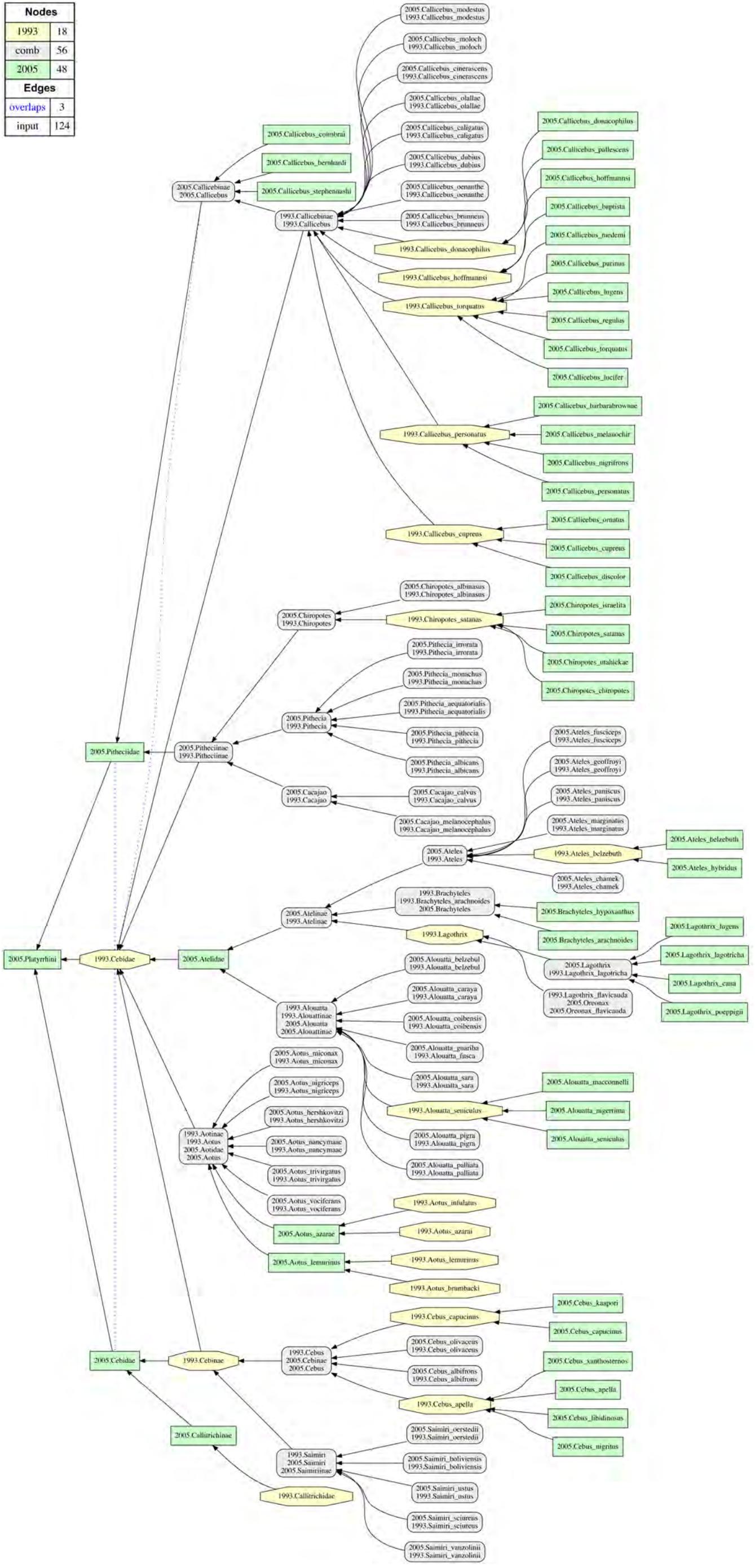

# Fig. S3-7

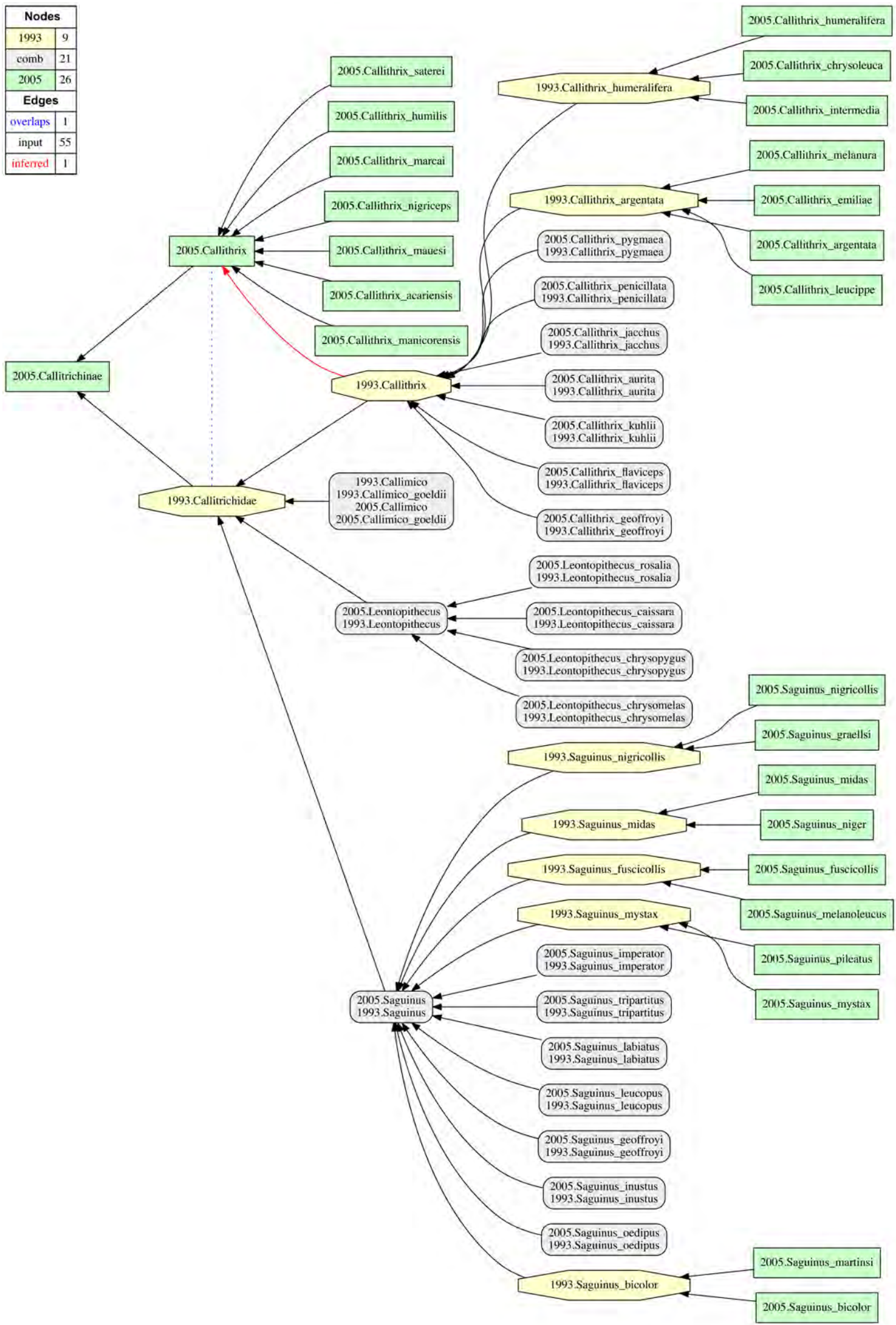

# Fig. S3-8

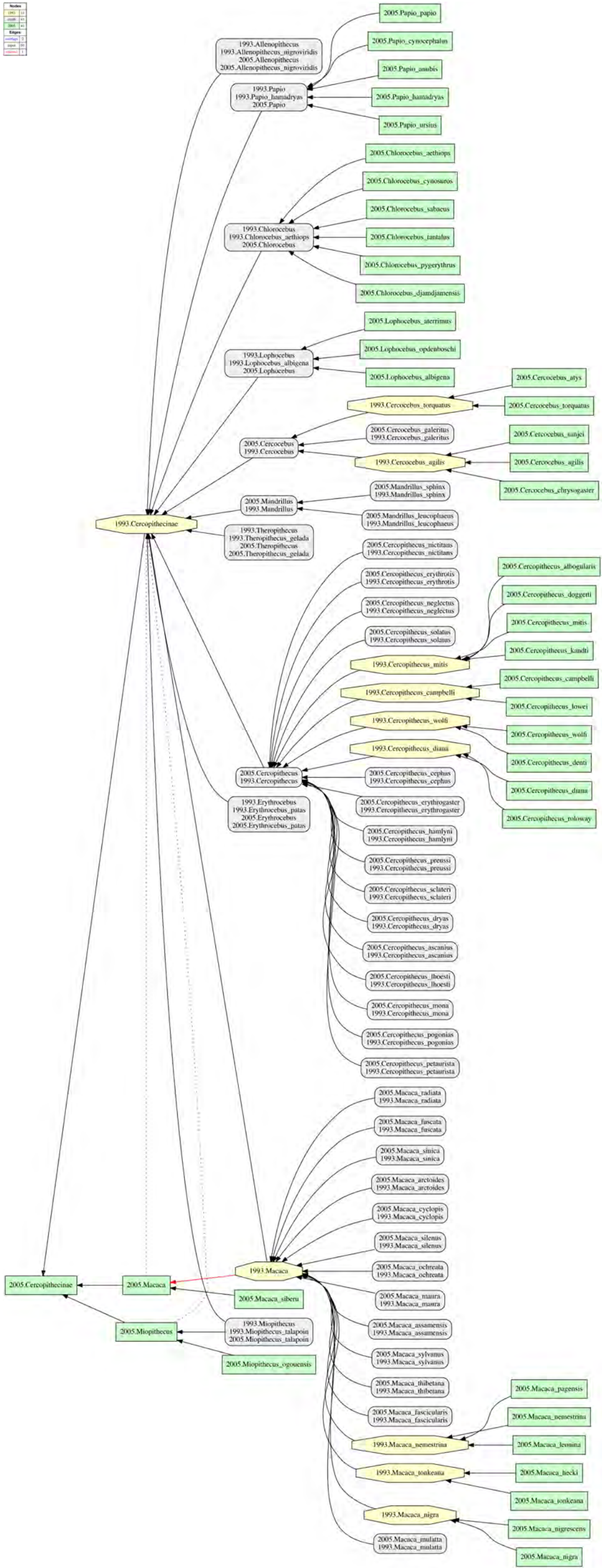

# Fig. S3-9

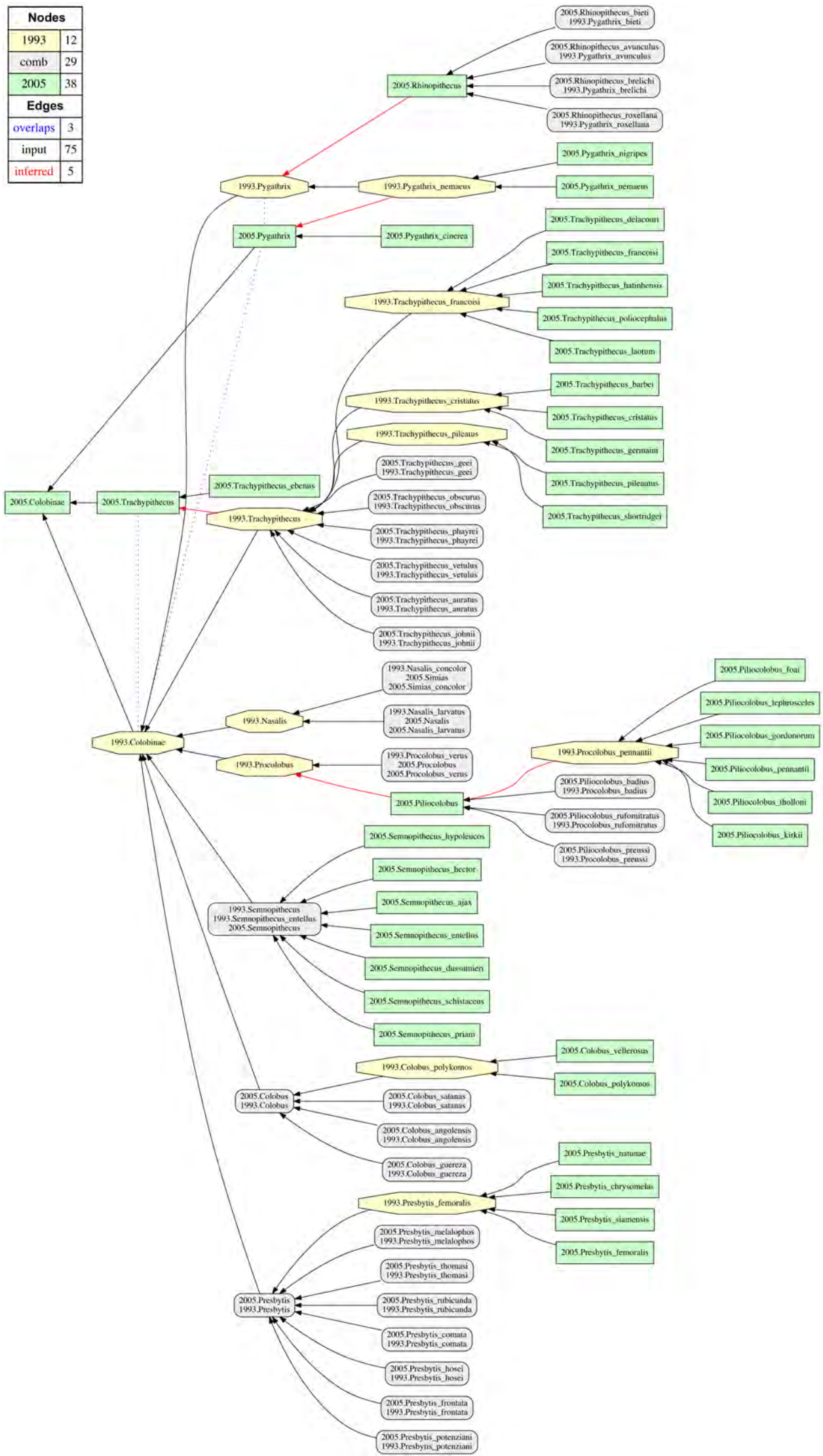

# Fig. S3-10

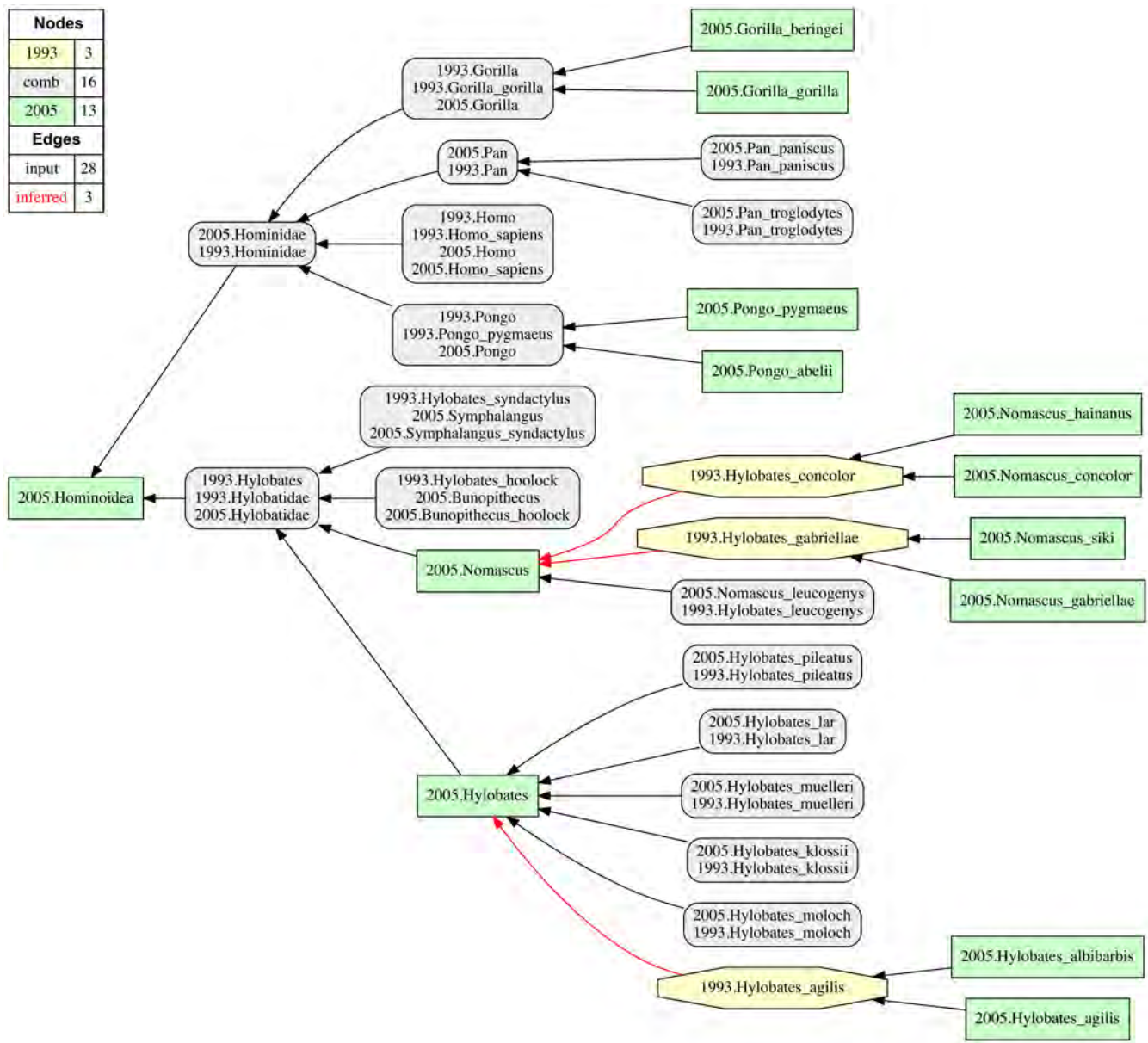